\documentclass[pdflatex,sn-mathphys-num]{sn-jnl}% Math and Physical Sciences Numbered Reference Style
\usepackage{color}

\usepackage{graphicx}
\usepackage[ansinew]{inputenc} % set .tex file encoding to ANSI
\usepackage{array} % used for improved array and tabular environments
\usepackage{amsxtra}
\usepackage{amstext}
\usepackage{amsthm} % Necessary for proof environment
\usepackage{latexsym}
\usepackage{IEEEtrantools} % provides improved version of eqnarray, called IEEEeqnarray
\usepackage{verbatim} %Necessary for comment environment
\usepackage{bm} % provides \bm command for bold math symbols
\usepackage{hyperref} % enables intra-pdf hyperlinks and PDF bookma
\usepackage{multirow}%
\usepackage{amsmath,amssymb,amsfonts}%
\usepackage{amsthm}%
\usepackage{mathrsfs}%
\usepackage[title]{appendix}%
\usepackage{xcolor}%
\usepackage{textcomp}%
\usepackage{manyfoot}%
\usepackage{booktabs}%
\usepackage{algorithm}%
\usepackage{algorithmicx}%
\usepackage{algpseudocode}%
\usepackage{listings}%

\newcommand\revisionC[1]{\textcolor{black}{#1}}

\begin{document}

\title[Article Title]{  Universal topology of exceptional points in nonlinear non-Hermitian systems}

%\author*[1,2]{\fnm{First} \sur{Author}}\email{iauthor@gmail.com}

%\affil*[1]{\orgdiv{Department}, \orgname{Organization}, \orgaddress{\street{Street}, \city{City}, \postcode{100190}, \state{State}, \country{Country}}}

\author[1]{\fnm{N. H.}  \sur{Kwong}}\email{kwong@optics.arizona.edu}
\affil*[1]{\orgdiv{College of Optical Sciences}, \orgname{University of Arizona}, \orgaddress{\street{1630 E. University Blvd.}, \city{Tucson}, \postcode{85721}, \state{Arizona}, \country{USA}}}

%\affiliation{Wyant College of Optical Sciences, University of Arizona, Tucson, AZ 85721}

\author[2,3]{\fnm{Jan}\sur{ Wingenbach}}\email{jawi1@mail.uni-paderborn.de}

\affil*[2]{\orgdiv{Department of Physics and Center for Optoelectronics and Photonics Paderborn (CeOPP)}, \orgname{Paderborn University}, \orgaddress{\street{Warburger Str. 100}, \city{Paderborn}, \postcode{33098},  \country{Germany}}}

\affil[3]{\orgdiv{Institute for Photonic Quantum Systems (PhoQS)}, \orgname{Paderborn University}, \orgaddress{\street{Warburger Str. 100}, \city{Paderborn}, \postcode{33098},  \country{Germany}}}

%\affiliation{Department of Physics and Center for Optoelectronics and Photonics Paderborn (CeOPP), Paderborn University, 33098 Paderborn, Germany}
%\affiliation{Institute for Photonic Quantum Systems (PhoQS),
%Paderborn University, 33098 Paderborn, Germany}

\author[2,3]{\fnm{Laura} \sur{Ares}}\email{laura.ares.santos@uni-paderborn.de}

%\affiliation{Department of Physics and Center for Optoelectronics and Photonics Paderborn (CeOPP), Paderborn University, 33098 Paderborn, Germany}
%\affiliation{Institute for Photonic Quantum Systems (PhoQS),
%Paderborn University, 33098 Paderborn, Germany}

\author[2,3]{\fnm{Jan} \sur{Sperling}}\email{jan.sperling@uni-paderborn.de}

%\affiliation{Department of Physics and Center for Optoelectronics and Photonics Paderborn (CeOPP), Paderborn University, 33098 Paderborn, Germany}
%\affiliation{Institute for Photonic Quantum Systems (PhoQS),
%Paderborn University, 33098 Paderborn, Germany}

\author[2]{\fnm{Xuekai} \sur{Ma}}\email{xuekai.ma@gmail.com}
%\email{xuekaima@mail.uni-paderborn.de}

%\affiliation{Department of Physics and Center for Optoelectronics and Photonics Paderborn (CeOPP), Paderborn University, 33098 Paderborn, Germany}

\author*[1,2,3]{\fnm{Stefan} \sur{Schumacher}}\email{stefan.schumacher@upb.de}
%\email{stefan.schumacher@uni-paderborn.de}

%\affiliation{Wyant College of Optical Sciences, University of Arizona, Tucson, AZ 85721}
%\affiliation{Department of Physics and Center for Optoelectronics and Photonics Paderborn (CeOPP), Paderborn University, 33098 Paderborn, Germany}
%\affiliation{Institute for Photonic Quantum Systems (PhoQS),
%Paderborn University, 33098 Paderborn, Germany}

\author*[1,4]{\fnm{R.} \sur{Binder}}\email{binder@optics.arizona.edu}
\affil[4]{\orgdiv{Department of Physics}, \orgname{University of Arizona}, \orgaddress{\street{1630 E. University Blvd.}, \city{Tucson}, \postcode{85721}, \state{Arizona}, \country{USA}}}

%\affiliation{Wyant College of Optical Sciences, University of Arizona, Tucson, AZ 85721}
%\affiliation{Department of Physics, University of Arizona, Tucson, AZ 85721}

%\pacs{???}

\date{\today}

%======================================================
%   Abstract   Nat. Comm. limit 150 words
%
\abstract{
Exceptional points (EPs)
are non-Hermitian degeneracies where eigenvalues and eigenvectors coalesce, 
giving rise to unusual physical effects across scientific disciplines. The concept of EPs has recently been extended to nonlinear physical systems.  We theoretically demonstrate a 
universal topology
%\revisionCB{universal topology, defined here as the topology of a universal unfolding,}
in the nonlinear parameter space for a large class of physical systems that support $2^\mathrm{nd}$ order  EPs in the linear regime. Knowledge of this topology (called elliptic umbilic singularity in bifurcation theory) deepens our understanding of $2^\mathrm{nd}$ order linear EPs, which here emerge as coalescence of 4 nonlinear eigenvectors. This helps guide future experimental discovery of nonlinear EPs and their classification, 
establish rigorous bounds of sensitivity enhancement of EPs in nonlinear systems,
and helps envision and optimize technological applications of nonlinear EPs. 
   Our theoretical approach is general and can be extended to nonlinear perturbations of $3^\mathrm{rd}$ and higher-order EPs.
}

\maketitle

%\narrowtext
%\begin{multicols}{2}

   Exceptional points (EPs), which are spectral degeneracies of non-Hermitian systems at which both eigenvalues and eigenvectors coalesce (mathematical definitions and generalizations are in \cite{kato.1966,seyranian-etal.2005,ashida-etal.2021}), have been increasingly explored across a broad range of scientific areas, including anharmonic oscillators \cite{bender-wu.1969},
    diffraction optics \cite{berry-odell.1998},
    quantum chaos \cite{leyvraz-heiss.2005,heiss.2012},
    conventional lasers \cite{brandstetter-etal.2014,lietzer-etal.2012},
    waveguides and photonic systems \cite{kullig-etal.2023,longhi.2018,quirozjuarez.2019,dong-etal.2020},
    atomic gases, condensates and high-Q cavity atomic lasers \cite{xu-etal.2017,choi-etal.10,wang-etal.2024},
    polariton lasers, microlasers  and polariton condensates \cite{dembowski-etal.01,%
    miao-etal.2016,%
    gao-etal.15,%
    gao2018chiral,%
    ozturk-etal.2021,%
    khurgin.2020,%
    Li2022,%
    PhysRevLett.122.185301,%
    PhysRevB.104.235408,%
    Li2022},
    non-Hermitian Bose--Hubbard models \cite{graefe-etal.2008},
    critical fluctuations and fluctuation spectra \cite{hanai-littlewood.20,binder-kwong.2021},
    and sensing applications \cite{wiersig.2014,hodaei-etal.2017,langbein.2018,wiersig.2020,sakhdari-etal.2019,zheng-chong.2025},
    non-Hermitian parity-time (PT) symmetric quantum systems
    \cite{%
    bender-hook.2024,%
    hassan-etal.2015,%
    miri-alu.2016,%
    ge-elganainy.2016,%
    kominis-etal.2017,%
    teimourpour-etal.2017,%
    ju-etal.2019,%
    xia-etal.2021,%
    lee-etal.2024,%
    klauck-etal.2025},
    and
    quantum state control  \revisionC{and squeezing \cite{abbasi-etal.2022,blinova-etal.2024}.}
    EPs have also been studied in the context of topological insulators and other topological systems (e.g.\
    \cite{luitz-etal.2019,%
    kawabata-etal.2019,%
    xiao-etal.2020,%
    tang-etal.2020,%
    bergholtz-etal.2021,%
    hu-etal.2022,%
    yokomizo-etal.2020,%
    ding-etal.2022,%
    jia-etal.2023,%
    lai-etal.2024}).

    While in most of these physical realizations EPs emerge as eigenvalues $\lambda $ of a linear non-Hermitian matrix $M$, i.e.\ from the eigenvalue problem  $M \textbf{x} = \lambda \textbf{x}$, more recent studies of EPs include nonlinear physical systems, such as \revisionC{optical or electric resonators,
    \cite{suntharalingam-etal.2023},}
    models of population dynamics \cite{felski-etal.2025}, as well as polariton condensates.
    In these systems, the complex interplay of nonlinear and non-Hermitian physics leads to counterintuitive phenomena such as shifts \cite{PhysRevA.103.043510,el2023tracking,Ramezanpour:24,wingenbach-etal.2024} and rotations \cite{wingenbach-etal.2024} of  the EP in parameter space through blueshifts and saturable gain.
    Moreover, mode switching can be realized in bistable regions near the EP \cite{wang2019dynamics}.
    In  Refs. \cite{wingenbach-etal.2024,Li2024}, it was shown that nonlinearity can alter the eigenvalue splitting in the vicinity of an EP and highly sensitive EP-based sensors have been realized in nonlinear non-Hermitian systems \cite{PhysRevApplied.18.054059,Bai-etal.2023,bai-etal.2024}. 
    \revisionC{
    The underlying eigenvalue problem here may be nonlinear in the sense that the matrix $M$ may depend on the eigenvalue $\lambda$ 
    (e.g. \cite{isobe-etal.2024}), the eigenvector $\textbf{x}$ (e.g. \cite{sone-etal.2022}), or both. In some instances, the nonlinear dependence on the eigenvector can be interchanged by that on the eigenvalue, e.g. \cite{zheng-chong.2025}. Other kinds of nonlinear eigenvalue problems are discussed in \cite{bender.2017}.
    In the following, we concentrate on the case where the nonlinear eigenvalue problem is formulated as a matrix equation where the matrix dependends on the eigenvector.
}
    For example, the energies in an optical microcavity,  which are described by the diagonal elements of $M$, can shift with the intensity of the light field described by $\textbf{x}$, hence $ M(\textbf{x}) \textbf{x} = \lambda \textbf{x} $.
    In the following, we refer to EPs that are solutions of a linear eigenvalue problem as `linear EPs,' and those solving a nonlinear eigenvalue problem are called `nonlinear EPs.'

   The wide range of physical systems in which EPs can occur raises the question  whether there exist any guiding principles that help clarify the occurrence of EPs as the physical parameters
    (in particular the nonlinearity)
   entering the matrix $M$ 

   are varied.%   \sout{for example, the  mode coupling in classical diffraction gratings \cite{berry-odell.1998} or frequency and coupling constants of ring resonators in sensing applications %\cite{wiersig.2020}).}

    In this paper, we demonstrate that 
   important
    insight can be obtained on the distribution of EPs in parameter space, especially for nonlinear eigenvalue problems. We show that a powerful way to do this is to consider the behavior of eigenvectors around EPs as a bifurcation problem and make use of established mathematical results in catastrophe theory.

\begin{figure}[t]
      \includegraphics[width=.5 \textwidth]{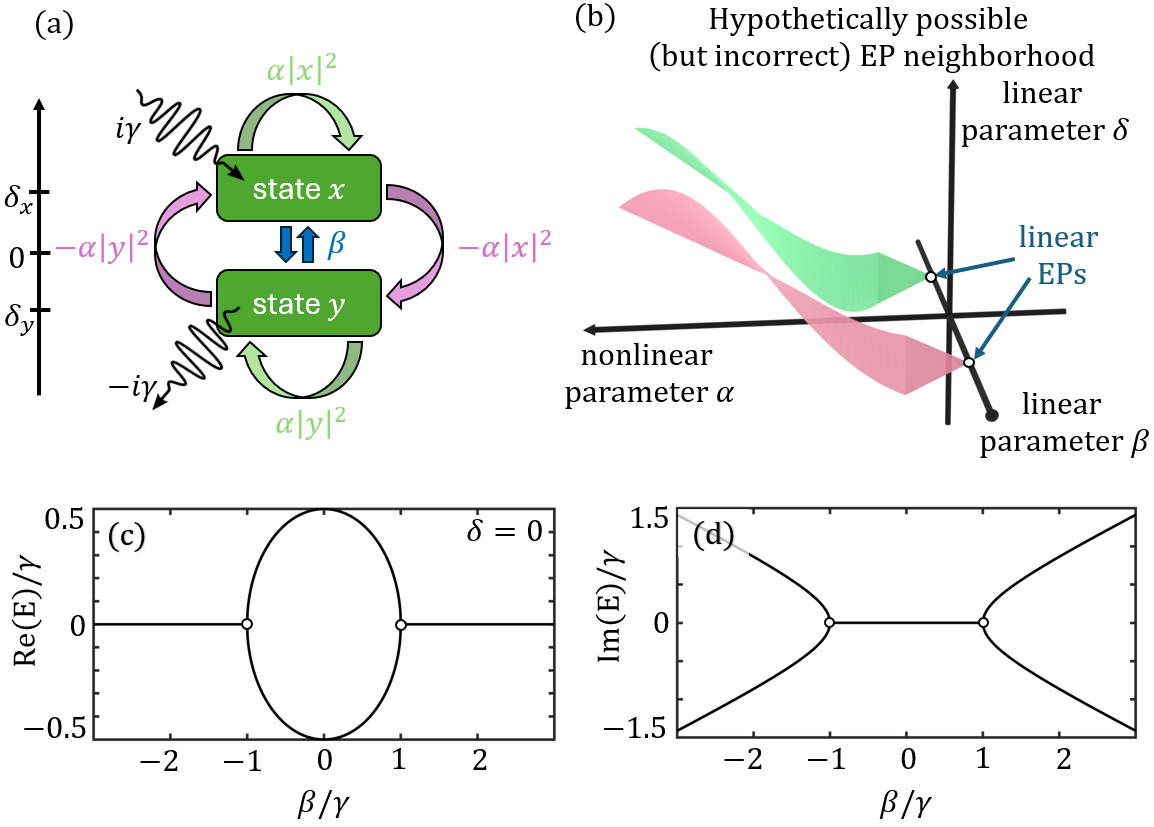} %0.8
    %\begin{center}
    \caption{%
    {\bf \revisionC{ Non-Hermitian two-state system.} }
        (a) Sketch of a non-Hermitian and nonlinear two-state (or two-mode) system, with linear coupling $\beta$, energy splitting $2\delta=\delta_x-\delta_y$, and loss/gain difference $2\gamma$.
        (b) Sketch of hypothetical location of exceptional points as a function of physical parameters.
        Without a nonlinearity, $\alpha=0$, isolated linear EPs are at $( \beta, \delta) = ( \pm \gamma, 0)$;
        in the nonlinear regime, $\alpha \neq 0$, EPs are implied on the colored surfaces.
        To illustrate the problem our study is solving, the surfaces we show here are one of infinitely many hypothetically possible (but generally wrong) EP neighborhoods.
        The correct geometric shape is shown in Fig. \protect\ref{fig:3D-elliptic-umbilic}.
        Panels (c) and (d) show the eigenvalues in the linear case ($\alpha = 0$) with two isolated  EPs at $(\beta,\delta) = (\pm \gamma, 0)$.}
    \label{fig:sketch-incorrect-neighborhood}
    %\end{center}
\end{figure}

    We illustrate this rigorous procedure with an analysis of nonlinear parameter-dependent $2 \times 2$ matrices,
   schematically shown in Fig. \ref{fig:sketch-incorrect-neighborhood},
    and show that a broad class of 
    physically different nonlinear systems share the same EP neighborhood geometry that fixes one (and only one) topology, namely a cone-like structure
    (with deltoid cross section) that consists of three fold-bifurcation surfaces attached to each other at cusp-bifurcation ribs.
    As a corollary, we find the possible numbers of solutions (eigenvalues) in each region of parameter space.
All these insights are based on one single equation (namely that for the so-called Lyapunov potential).
   Our findings can be used in future studies to establish rigorous bounds on the sensitivity enhancement of EPs  in nonlinear systems.
   It also provides guardrails against possibly wrong conclusions, illustrated in Fig. \ref{fig:sketch-incorrect-neighborhood}(b),
   that might be obtained from 
     limited experimental observations or approximate theories.

    To solve the highly sophisticated problem outlined above, we utilize the general theorems of a branch of bifurcation theory called catastrophe theory
    \cite{%
    zeeman.1976b,%
    thom.1989,%
    arnold.1992,%
    saunders.80,%
    gilmore.81,%
    poston-stewart.1978}.
    Roughly speaking, the framework of catastrophe theory describes the shapes of singularity sets -- or phase boundaries in nonequilibrium phase diagrams, to use a physicist's language -- related to singularities of potential functions $V_L$ (Lyapunov potentials).
    The word catastrophe is used for the sudden qualitative change or bifurcation at a `phase boundary.'
    Moreover, it lists all possible shapes (`elementary catastrophes') for cases in which the number of variables (often called state variables) and the number of control parameters (i.e. external parameters) are relatively small,
    see Supplementary Table 1 of the Supplementary Information.
   % ref{catastrophes.tab}

    Examples of systems that have been analyzed by catastrophe theory include dynamical systems obeying equations of motion of the form
    $\dot{\textbf{x}} = - {\rm grad} \{ V_L( \textbf{x} )\}  $, where the phase boundaries in the control parameter space follow from the assumption of degenerate stationary states (degenerate critical points), corresponding to $  {\rm grad} \{ V_L( \textbf{x} ) \} = 0$ and $  {\rm det} \{ h \} = 0 $,  where $h$ denotes the Hessian matrix of $V_L$.
    Similarly, in optics, 
    \revisionC{the solution of the Helmholtz equation,
    which describes the stationary solutions of a second-order time-differential equation, in contrast to the example of
    a first-order time-differential equation mentioned above,
    can be evaluated 
    }
    with the method of stationary phase  and stationary points, as well as caustics occurring when two or more stationary points coalesce (p.\ 336 of \cite{adam.2002}).
    This identifies rainbows and other optical phenomena as diffraction catastrophe, specifically fold and cusp
    bifurcations; see \cite{berry-upstill.1980,berry.90,adam.2002} and Ch.\ 12 of \cite{poston-stewart.1978} for details.
    Higher bifurcations (catastrophes), such as the  so-called elliptic umbilic, have also been studied in the context of optical diffraction \cite{berry-etal.79,berry.90}.
    Many other areas of application of catastrophe theory can be found in
    \cite{%
    zeeman.1976a,%
    zeeman.1976b,%
    saunders.80,%
    gilmore.81,%
    rosen.1988,%
    poston-stewart.1978,%
    heiss.04,%
    longhi.2018,%
    heiss.2012,%
    barkleyrosser.2007}.
    This further extends to studies in biology, social sciences, and economics, with various degrees of acceptance because low-dimensional dynamical-systems models are not always appropriate for the simulation of complex systems in those sciences \cite{rosen.1988,barkleyrosser.2007}.
    In contrast, in many areas of physics low-dimensional models and eigenvalue problems (e.g. two--level models) are well established.
    %}

\begin{figure}[t]
    \includegraphics[width=0.5 \textwidth]{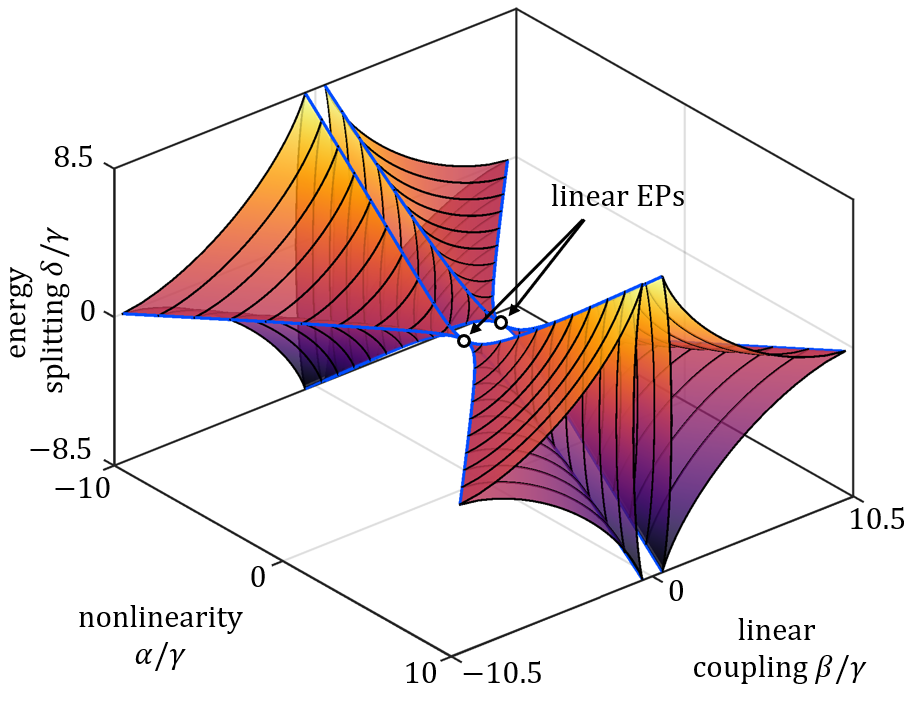} %0.8
    %\begin{center}
    \caption{%
     {\bf \revisionC{ Elliptic umbilic singularity set.} }
        Surface of EP locations as a function of physical parameters (singularity set)
        explained in Fig.~\protect\ref{fig:sketch-incorrect-neighborhood}.
        In the weakly nonlinear regime, $\alpha \ll \gamma$, the shape approaches that of an  elliptic umbilic
        bifurcation
        (`catastrophe').
        The shape of each singularity set
        consists of two infinite three--cusped conical surfaces, the apices of which meet at the two linear EPs $(\alpha, \beta,\delta) = (0, \pm \gamma, 0)$.
        At large $\alpha$, the three--cusped shape remains but is deformed guaranteeing that the two cone-like surfaces originating at $\beta= \pm 1$ do not overlap;
a        see Fig.~\ref{fig:elliptic-umbilic-cross-section}.}
    \label{fig:3D-elliptic-umbilic}
    %\end{center}
\end{figure}

\begin{figure}[t]
    \includegraphics[width=.45 \textwidth]{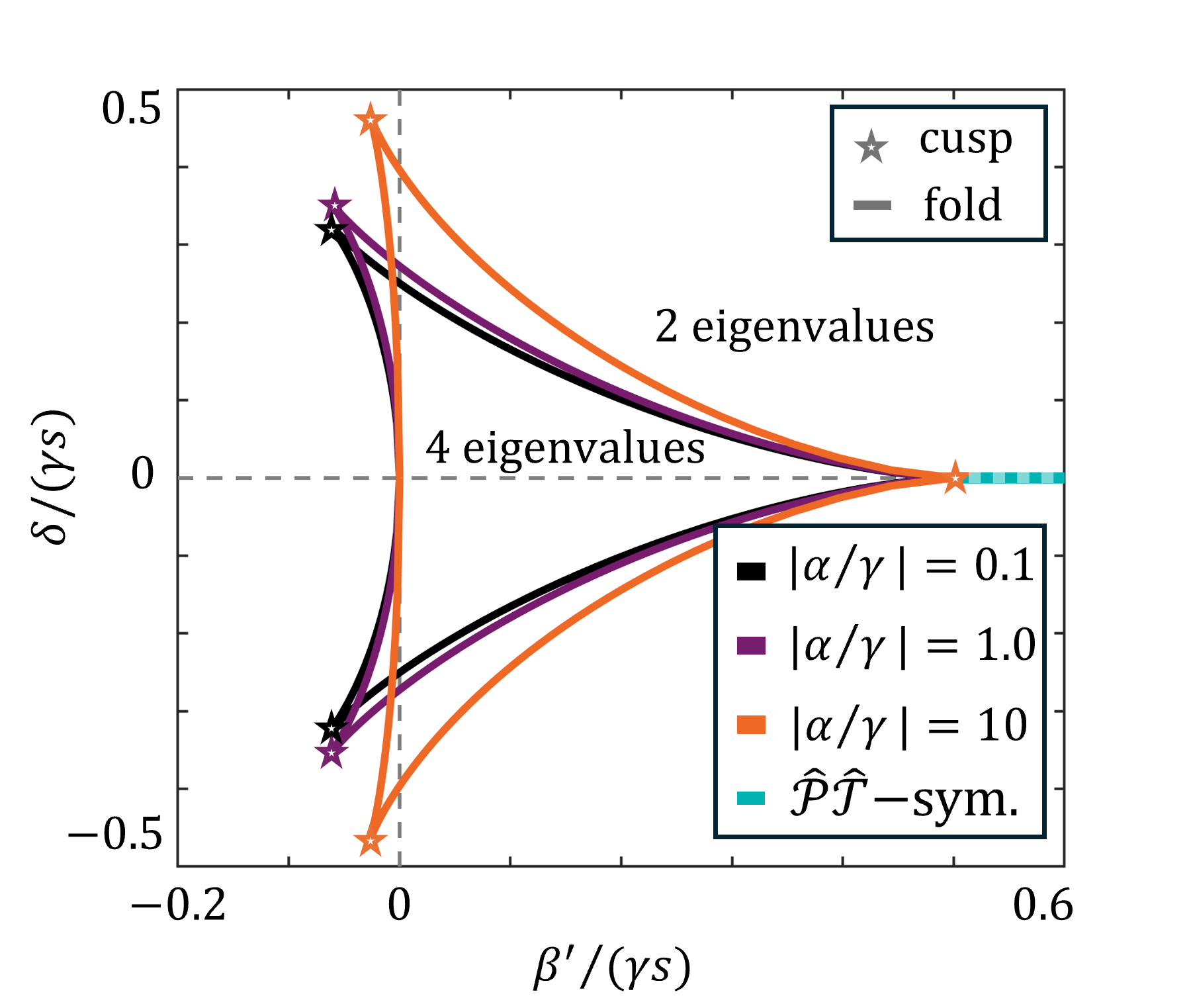}
    %\begin{center}
    \caption{%
     {\bf \revisionC{ Cross section of singularity set.} }
        Cross section of EP surface for fixed nonlinearity $\alpha$;
        axis scaled with scale factor $s=2( \sqrt{1+ \alpha^2 / \gamma^2} - 1 )$. Here, $\beta' = \beta - \gamma$.
        Small $\alpha \ll \gamma$ (vicinity of linear EP, Eq. \ref{EU-1.equ}):
        deltoid or three-cusped hypocycloid \protect\cite{broecker-lander.1975,poston-stewart.1978,saunders.80} of the elliptic umbilic surface with exact scaling $s= \alpha^2 / \gamma^2 $.
        Large $\alpha$ (farther away from linear EP, Eq. \ref{Lyapunov.equ}):
        exact scaling is lost, but three--cusped shape remains.
        The half-infinite line where PT symmetry is unbroken is indicated by the blue line.
    }
    \label{fig:elliptic-umbilic-cross-section}
    %\end{center}
\end{figure}

\revisionC{
In this study, we aim to understand the topological structure of the EP set around the linear EP in parameter space when nonlinearity in the matrix is turned on. Our strategy of utilizing catastrophe theory is as follows 
(more details are given in Methods and Sec. II of the Supplementary Information). The layout of the catastrophe theory formulation consists of a state variable space and a control parameter space. The eigenvector elements of the nonlinear matrix are considered as values of state variables determined by the eigenvector equations, and the coordinates of the control parameter space are the parameters in the matrix.  A Lyapunov potential function is constructed under the condition that its critical points (in state variable space) are equal to the eigenvector elements. The coalescing eigenvectors are accordingly given by the degenerate critical points of the Lyapunov potential. The set of degenerate critical points forms a hyper-surface in the combined state variable space and parameter space. The projection of this hyper-surface onto the parameter space yields the set of EPs in parameter space. 
}

\revisionC{
To obtain a topological characterization of the nonlinear EP set in parameter space around the linear EP, we Taylor expand the Lyapunov potential in both the state variables and the parameters about this point. This allows us to match our Lyapunov potential locally with one of the known catastrophes. 
See, for example, p. 137 of \cite{saunders.80} and p. 11 of \cite{gilmore.81} for a list of the `elementary catastrophes', which is also reproduced as Supplementary Table I in the Supplementary Information. We find that, for the broad class of nonlinear $2 \times 2$ matrices considered here, the matched catastrophe is the elliptic umbilic. Consequently, the EP sets (in parameter space) of these matrices share a common topological structure which is that of the set of degenerate critical points of this catastrophe. 
}

\revisionC{
 \section{Results}   \label{Sec:results}
}

\revisionC{
    In the following, we present our approach for a large class of $2 \times 2$ matrices and analyze the EP structure in parameter space.
    We find that, for all matrices in this class, the linear EPs are the organizing points of an elliptic umbilic singularity.
    We stress that, in addition to the matrices discussed below, there may be other $2 \times 2$ matrices with elliptic umbilic nonlinear EP structures that we have not yet considered. 
    }

\revisionC{
    In the next  subsection, Sec. \ref{Sec:quadratic-nonlinearity}, we illustrate our findings using a specific example, namely a symmetric matrix with Kerr-like nonlinearity, and in Sec. \ref{Sec:extension-beyond-quadr-nonlin}-\ref{Sec:ecomplex-alpha}
    we  discuss extensions and generalizations of that example. All examples discussed below belong to the class of matrices that can be written in the following form
     (now using the symbol $H$ as commonly used in the context of non-Hermitian Hamiltonians)
  %  \begin{widetext}
\begin{equation}\label{equ:Hnonlinear-cubic_1}
H =
\begin{pmatrix}
      \delta + i \gamma     + \alpha F(w)  & b \\
c  & - \delta - i \gamma  - \alpha F(w)
\end{pmatrix}\,
\end{equation}
%  \end{widetext}
with 
$\delta, \gamma, \in\mathbb R$,
$b, c, \alpha \in\mathbb C$,
and the nonlinearity
$F(w)$  
being a real-valued analytic function of the eigenvector parameter $w$ in the vicinity of the linear EP. Specifically, the real-valued variable $w$ is related to the (right) eigenvector of $H$,
   which, normalized to $1$, is written as $[ x \,\, y ]^T = (1 / \textit{A} ) [ 1 \,\, \tilde{w} ]^T$, $\tilde{w} = w e^{i \theta}$, $\textit{A} = \sqrt{1 + w^2}$. 
   The matrix in Eq.~(\ref{equ:Hnonlinear-cubic_1}) is a useful starting point for our analysis. The fact that $H$ is traceless, in other words that the average of the diagonal elements is zero, is not a restriction. 
   Below, in Sec. \ref{Sec:extension-beyond-quadr-nonlin}, we will discuss in more detail the procedure of how to remove the average of the diagonal elements without changing the eigenvectors. 
   }

\revisionC{
 \subsection{Example: quadratic nonlinearity}  \label{Sec:quadratic-nonlinearity}
}

\revisionC{In order to give a concrete example, 
    in this section we illustrate our approach for symmetric $2 \times 2$ matrices with a second--order linear EP
    and a nonlinearity that is a quadratic function of the eigenvector variable $w$ defined above (Kerr-like nonlinearity). Furthermore, we assume that at the EP, where $w=1$, the lowest-order Taylor expansion of the nonlinearity function yields  $F(1) = 0$    and  $F'(1) \neq 0 $.
     }  We write the nonlinear eigenvalue problem
    as
    \begin{equation}
        \label{eigen-1.equ}
        H
        \begin{bmatrix}
        x \\ y
        \end{bmatrix} =
        \begin{bmatrix}
        a & \beta \\
        \beta &  - a
        \end{bmatrix}
        \begin{bmatrix}
        x \\ y
        \end{bmatrix} =
        E
        \begin{bmatrix}
        x \\ y
        \end{bmatrix}\,,
    \end{equation}
    where $a = \delta + i \gamma + \alpha \left( | x |^2 - | y |^2  \right) $, $\alpha, \beta, \gamma, \delta\in\mathbb R$ are parameters, and all variables and parameters are assumed scaled and unitless.
   This parametric form is chosen for ease of analysis and the specific form of the nonlinearity includes a broad range of physical systems, for example photonic systems with Kerr-type nonlinearity, atomic and polariton condensates, and superfluids \cite{Liu2000,berloff1999,Kivshar1989,Dominici2015}; 
   %More generally, the restrictions to a symmetric matrix and to a quadratic nonlinearity (in the matrix element) 
    the restrictions can be relaxed and the method is still applicable
   (see \revisionC{ Sec. \ref{Sec:extension-beyond-quadr-nonlin} and}
   Sec. I of the Supplementary Information for details).
    In the linear limit, the two states  
   % basis  channels
    are separated by $2 \delta$ in frequency, and one state gains and the other dissipates at equal rates
    $| \gamma |$.
    %
 %   The eigenvector, normalized to $1$, is written as $[ x \,\, y ]^T = (1 / \textit{A} ) [ 1 \,\, \tilde{w} ]^T$, $\tilde{w} = w e^{i \theta}$, $\textit{A} = \sqrt{1 + w^2}$.
    
    Eliminating the eigenvalue $E$ from Eq.~\eqref{eigen-1.equ}  reduces the matrix equation to a nonlinear equation in the complex eigenvector element $\tilde{w}$, the real and imaginary parts of which are
    \begin{align}
        \label{eigen-2-real.equ}
        2 \alpha \left[ \frac {1 - w^2} {1 + w^2} \right] + 2 \delta + \beta \left[ w - \frac {1} {w} \right] \cos \theta &= 0 \,, \\
        2 \gamma + \beta \left[ w + \frac {1} {w} \right] \sin \theta &= 0 \,. \label{eigen-2-imag.equ}
    \end{align}
    Formulating the eigenvalue problem this way enables the map to the function singularity problem to be constructed.
    In this context, we call $(w , \theta)$ the state variables and $(\alpha, \beta, \gamma, \delta)$ the control parameters.
    The first step in the map is to construct a Lyapunov function $V_L (w, \theta, \alpha, \beta, \gamma, \delta)$ such that its critical point equations in state variable space,  $\partial V_L / \partial w = 0$, $\partial V_L  /  \partial \theta = 0$, are the same as the eigenvector equations \eqref{eigen-2-real.equ} and \eqref{eigen-2-imag.equ}.
    It can be directly verified that the Lyapunov function
    \begin{eqnarray}
        \label{Lyapunov.equ}
        V_L (w, \theta, \alpha, \beta, \gamma, \delta) & =  & - 2 \alpha \ln \left[ \frac {g} {2} \right] + 2 \delta \ln w   \nonumber  \\
        & &+ \beta g \cos \theta - 2 \gamma \left[ \theta - \frac {3 \pi} {2} \right] \,,
    \end{eqnarray}
    with $g=w + w^{-1}$, satisfies this requirement.

    To see the connection with catastrophe theory, we need to consider $V_L$ in the vicinity of the linear EPs.
    Since $V_L$ only has terms that are linear in the control parameters, we can reduce the dimensions of the parameter space by rescaling by one of the parameters, which we choose to be $\gamma$.
    The parameter space for the EP analysis, $( \alpha / \gamma , \beta / \gamma , \delta / \gamma )$, is thus three-dimensional.
    In the linear limit ($\alpha / \gamma = 0$), the matrix $H$ has two separate EPs (bifurcation points) in the $( \beta / \gamma , \delta / \gamma)$ plane at $(\beta / \gamma , \delta / \gamma ) = (\pm 1 , 0)$.
    The coalescing eigenvector element (degenerate critical point) is $(w , \theta) = (1 , 3 \pi / 2)$ for $\beta / \gamma = 1$ and $(w , \theta) = (1 , \pi / 2)$ for $\beta / \gamma = -1$.
    Since the critical point behavior around the two linear EPs is related by reflection symmetry about the plane $\beta = 0$, it suffices to analyze one, $\beta / \gamma = 1$, of the two.
    We choose, for convenience, the $(w , \theta)$-independent terms in Eq. \eqref{Lyapunov.equ} for $V_L$ to shift $V_L $ to zero at the $\beta / \gamma = 1$ linear EP.
    We firstly consider the properties of the nonlinear EP set in the immediate neighborhood of this linear EP.
    We write $r = w - 1$, $\phi = \theta - 3 \pi / 2$, $\beta^\prime / \gamma = \beta / \gamma - 1$ and denote by $V_{EU} (r, \phi, \alpha / \gamma , \beta^\prime / \gamma , \delta / \gamma)$ the lowest-order terms in an expansion in $(r , \phi , \alpha / \gamma , \beta^\prime / \gamma, \delta / \gamma)$ of $V_L$.
    Then, we find
    \begin{equation}
        \label{EU-1.equ}
    \begin{aligned}
        {}& V_{EU} (r, \phi, \alpha / \gamma , \beta^\prime / \gamma , \delta / \gamma)
        \\
        ={}& -\left[ \frac {\phi^3} {3} - \phi r^2 + \frac {\alpha} {\gamma} r^2    -  2 \frac {\beta^\prime} {\gamma} \phi - 2 \frac {\delta} {\gamma} r \right]\,.
    \end{aligned}
    \end{equation}
    The form of Eq.~\eqref{EU-1.equ} matches that of a universal unfolding of the $D_{-4}$ singularity, also called elliptic umbilic singularity,
 %   catastrophe,
    around the linear EP.
    Its degenerate critical points and bifurcations, which are well understood, are obtained as simultaneous solutions to three equations:
    $\partial V_{EU} / \partial r = 0$, $\partial V_{EU} / \partial \theta = 0$,  and $\det \{ h \} = 0$, where $h$ is the Hessian matrix of $V_{EU}$.
%insert new sentences xxxxxx
\revisionC{
The corresponding set of three equations, with $V_L$ taking the place of  $V_{EU}$, give the EP set in the whole parameter space (the singularity set). 
Since there are five unknowns, $w$, $\theta$ ,    $\alpha/\gamma$,    $\beta' / \gamma$,     $\delta/\gamma$, the solution set of the three equations forms a 2D surface in the 5D
 combined state variable and parameter space. The projection of this solution surface onto the parameter space is the singularity set. Algebraic detail of the solution set can be found in Sec. III of the Supplementary Information  . 
}

    The singularity set, plotted in Fig. \ref{fig:3D-elliptic-umbilic}, consists of a pair of two conical surfaces with deltoid cross sections, the apices of which meet at the linear EPs at $(\alpha, \beta^\prime,\delta) = (0, 0, 0)$ and $(0, -2 \gamma, 0)$. There are four and two critical points in state variable space at each control parameter point inside and outside the cones, respectively;
    also see Fig. \ref{fig:elliptic-umbilic-cross-section} below.
    The ribs of each cone are cusp points where three critical points inside the cone meet and emerge as one critical point outside the cone.
    Each point on the surface between two ribs is a fold point where two critical points inside the cone meet and annihilate each other, thus vanishing on the outside.
    This arrangement of the fold and cusp singularities is characteristic of the elliptic umbilic catastrophe ($D_{-4})$.
    We call the meeting point of the cones the `organizing point' of the catastrophe.
    We show in Fig.~\ref{fig:elliptic-umbilic-cross-section} that the cone-like topological structure is not limited to the immediate neighborhood of the linear EP, but remains valid even at large nonlinearities (large $\alpha$).
    \revisionC{More information on the EP set and the Lyapunov potential is given in Sec. IV and V of the Supplementary Information. }

\revisionC{Interestingly, we see that the non-trivial topology of the set of EPs in the nonlinear parameters space (the parameter space that includes the nonlinearity $\alpha$, as shown in Fig. \ref{fig:3D-elliptic-umbilic}), is already fully determined by the EPs in the linear space (the subspace of parameters where the nonlinearity is zero). 
}
\revisionC{
In other words, there are infinitely many possibilities for the EP topology in the nonlinear case 
(cf. Fig. \ref{fig:sketch-incorrect-neighborhood}b), but already from the properties of the linear EP alone we find that the topology must be the  cone-like structure. At small nonlinearities, this structure is of the form of an
elliptic umbilic, i.e. a cone with a deltoid cross section (and the cone-like structure can persist at larger nonlinearities, as in our example, preserving the topology of the elliptic umbilic but not its precise geometry). 
}
\revisionC{
This is a remarkable statement, since usually one would try to find the set of EPs in the nonlinear parameters space by searching or scanning the part of the parameter space where the nonlinearity is non-zero.
To be more specific, if in Eq. \eqref{EU-1.equ} we set $\alpha=0$, $\beta'=0$ and $\delta=0$, we are left with the so-called `germ' (sometimes called degenerate core) of the catastrophe. For catastrophes with codimension up to 5 (called elementary catastrophes), the germs can be found in textbooks, e.g. \cite{sakhdari-etal.2019,gilmore.81} and are reproduced in Supplementary Table I. Each germ gives an unambiguous assignment of the corresponding elementary catastrophe. Since, in our case, the Lyapunov potential (as a polynomial in the state variables given by the eigenvectors) that generates the eigenvector equations of the nonlinear eigenvalue problem exists, we can predict the topology of the EP singularity set in the full parameter space solely from the Lyapunov potential where all parameters (including the nonlinearity) have been set to zero. 
The precise geometric shape can change under smooth transformations of parameters, but its topological properties are  determined by the universal unfolding (which is defined as the versal unfolding of minimal dimension, see, e.g., p. 621 of Ref. \cite{gilmore.81}) of the elliptic umbilic singularity. %It is in this sense that we refer to the cone-like elliptic umbilic EP structure as universal. 
}

    Regarding PT symmetry
    \cite{%
    hassan-etal.2015,%
    miri-alu.2016,%
    ge-elganainy.2016,%
    kominis-etal.2017,%
    teimourpour-etal.2017,%
    ju-etal.2019,%
    xia-etal.2021,%
    lee-etal.2024,%
    klauck-etal.2025,%
    bender-hook.2024},
    we  note that the Hamiltonian in Eq. \eqref{eigen-1.equ} is PT-symmetric, i.e. $[H,\mathrm{PT}]=0$, with parity represented by the $\sigma_1$ Pauli matrix, if $\delta=0$.
    In that case, PT symmetry is unbroken (i.e. the eigenenergies are real and the eigenvectors are common to $H$ and PT)  if $|\beta|  > \sqrt{\alpha^2 + \gamma^2}$.
    This region is indicated as thick blue line in Fig. \ref{fig:elliptic-umbilic-cross-section}.
\revisionC{
In Fig. \ref{fig:3D-elliptic-umbilic}, it is the part of the $\delta = 0$ plane outside of the deltoid cone.
}

\begin{figure}[t]
    \includegraphics[width=.5 \textwidth]{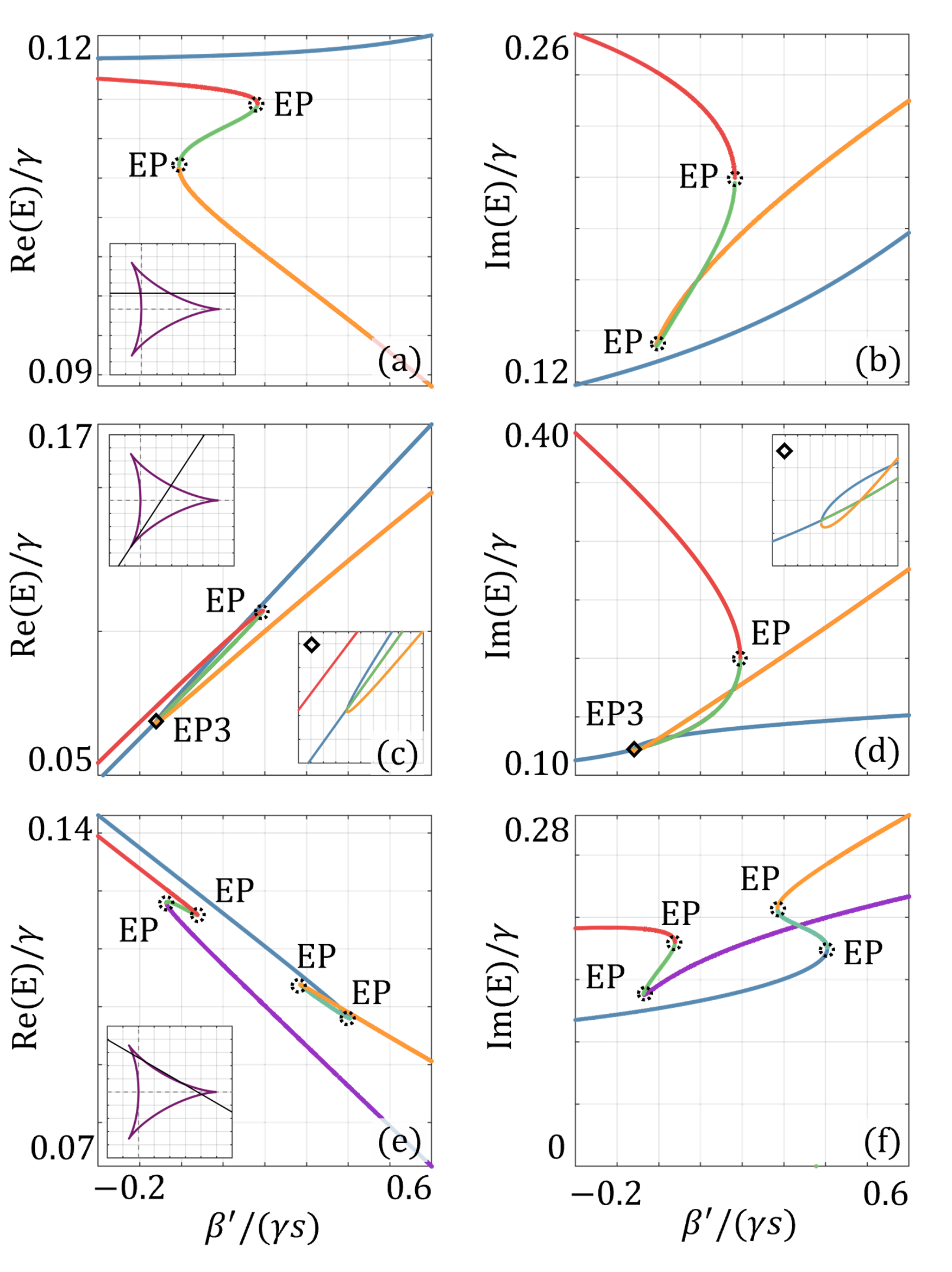}
    %\begin{center}
    \caption{%
      {\bf \revisionC{ Selected complex eigenvalue traces.} }
        Complex eigenvalues (real and imaginary parts) of $H$ in Eq \eqref{eigen-1.equ} at fixed, nonzero $\alpha/\gamma =1$ across lines in the $(\beta',\delta)$ plane shown as insets.
        Regardless of the lines' directions, we have two and four eigenvalues outside and inside the cone, respectively, in agreement with the number of values of critical points of the Lyapunov potential
       (see \revisionC{Supplementary Figure S3} for examples of the potential landscape).
        At the smooth parts of the deltoid boundary (fold lines), two states are created/annihilated, and two states are unaffected by the boundary (a,b,e,f).
        Crossing the cusp, (c,d), three states coalesce at a 3rd-order nonlinear EP, denoted as EP3, and one state is unaffected.
    }
    \label{fig:eigenvalues}
    %\end{center}
\end{figure}

    The power of using catastrophe theory in the context of EPs can be seen from the following considerations.
    First, the information on the geometric structure of the nonlinear EPs provides immediate knowledge of the number of eigenvalues as a function of the parameters used.
    The potential $V_{EU}$ has four critical points inside the cone and two outside the cone;
    exemplary plots of $V_{EU}$  are shown in Supplementary Figure S3 the Supplementary Information.
    This implies  that we have four eigenvalues inside the cone and two outside, and the number changes at the folds and cusps,
    which represent the location of the EPs.
    This is illustrated in Fig.~\ref{fig:eigenvalues}, which shows the eigenvalues of $H$ along the straight lines in ($\delta,\beta')$-parameter space as indicated in each row of subplots.
At the apices of the cones, Fig. \ref{fig:3D-elliptic-umbilic},
the four critical points coalesce, showing that, depending on how the limit is performed,
the linear EP can be viewed as the limiting case of a 4th-order nonlinear EP.
   A stability analysis of the solutions is planned as future work.

\revisionC{\subsection{Extension beyond quadratic nonlinearities}\label{Sec:extension-beyond-quadr-nonlin}}

\revisionC{
In the previous section, Sec. 
\ref{Sec:quadratic-nonlinearity}, we discussed the example of a Kerr-like  nonlinearity. We note, however, that}
%The second aspect highlighting 
the power of catastrophe theory lies in the fact that the linear matrix already fixes the elliptic umbilic catastrophe, and a  broad class of matrices extending our case given in Eq. \eqref{eigen-1.equ} will have the same \revisionC{topological} singularity structure.
\revisionC{
Let us first note that the matrix in Eq. \eqref{eigen-1.equ} is equivalent, as far as the topology of the EP object in parameter space is concerned, to 
one that contains self coupling ($\alpha_c$) and cross coupling ($\alpha_x$) terms with quadratic nonlinearities: 
%\begin{widetext}
\begin{equation}\label{equ:Hnonlinear-1}
H_1 =
\begin{pmatrix}
      \delta + i \gamma     + \alpha_c |x|^2    + \alpha_x |y|^2  & \beta \\
\beta  & - \delta - i \gamma  + \alpha_c |y|^2    + \alpha_x |x|^2
\end{pmatrix}\,.
\end{equation}
%\end{widetext}
Here, the parameter $\alpha$ used above corresponds to $\alpha = (\alpha_c - \alpha_x)/2$. 
The two matrices differ by a multiple of the unit matrix, and thus have the same eigenvectors while the 
eigenvalues differ only by a constant shift $\bar{\alpha} = (\alpha_c + \alpha_x)/2 $. Here, we assume the couplings $\alpha_c$ and $\alpha_x$ to be real-valued, but below we will see that they also can be imaginary (which in some applications can be related to gain saturation) or complex.
}
\revisionC{
We can generalize the matrix in Eq. \eqref{equ:Hnonlinear-1} to allow for sufficiently smooth real valued nonlinearity functions $f(u)$ (for which conditions will be given below) and with complex nonlinearity parameters $\alpha_c$, $\alpha_x$
%\begin{widetext}
%
%
\begin{equation}\label{equ:H1nonlinear-cubic}
H_1 =
\begin{pmatrix}
      \delta + i \gamma     + \alpha_c  f(|x|) +   \alpha_x  f(|y|)  & \beta \\
\beta & - \delta - i \gamma  - \alpha_x  f(|x|)   + \alpha_c f(|y|)
\end{pmatrix}\,.
\end{equation}
%
%
%
%
%\end{widetext}
%
%
Similar to the case discussed above, we can subtract a diagonal matrix 
that is a unit matrix multiplied by the (nonlinear) constant
$g(\mathrm{\bf x}) = \bar{\alpha} (f(|x|)    +  f(|y|))$, i.e. we define
$H_1 =g(\mathrm{\bf x}) \hat{1} + H $, where $\hat{1}$ is the unit matrix,
and obtain 
%\begin{widetext}
\begin{equation}\label{equ:Hnonlinear-cubic}
H =
\begin{pmatrix}
      \delta + i \gamma     + \alpha ( f(|x|)    -  f(|y|)  & \beta \\
\beta  & - \delta - i \gamma  - \alpha (f(|x|)    -  f(|y|))
\end{pmatrix}\,.
\end{equation}
%\end{widetext}
}
\revisionC{Once the nonlinear eigenvalue problem of $H$ has been solved and the eigenvalues $\lambda_n$ and eigenvectors $\mathrm{\bf x}_n$ are known, we evaluate the constant $g(\mathrm{\bf x})$ with the eigenvectors $\mathrm{\bf x}_n$ and obtain the eigenvalues of $H_1$ in Eq. (\ref{equ:H1nonlinear-cubic}) as
$\lambda_{n}^{(1)}$ = $\lambda_n + g(\mathrm{\bf x}_n)$, while the eigenvectors for $H_1$ and $H$ are the same.
As in the previous section, we write the eigenvector elements as functions of $w$ and $\theta$, and note that the moduli of both components,  $|x| = 1/\sqrt{1+w^2}$ and $|y| = w/\sqrt{1+w^2}$ , only depend on the real variable $w$.
 Hence the nonlinear term can be written as
\begin{equation}
\alpha ( f(|x|)    -  f(|y|) = \alpha F(w)
\end{equation}
with
\begin{equation}
F(w) = f( \frac{1}{\sqrt{1+w^2}}) - f( \frac{w}{\sqrt{1+w^2}})
\end{equation}
Similar to the example of the quadratic nonlinearity discussed in the previous section, the EP is at $w=1$.
As mentioned at the beginning of Sec. \ref{Sec:results}, the condition for the elliptic umbilic singularity is that at the linear EP (where $w=1$ in the case of symmetric matrices), the function $F(w)$
must be approximated by a power expansion.
Therefore, important examples of functions with elliptic umbilic singularity sets include all power functions
 $f(u)=u^n$ with nonzero exponent, $n \neq 0$, such as the quartic nonlinearity $f(u) = u^4$, and periodic functions such as $f(u) = \cos(u^2)$.
}

\revisionC{\subsection{Different Taylor expansions}\label{Sec:different-Taylor-expasions}}

\revisionC{
So far, we have limited the analysis to symmetric matrices where, at the EP (i.e. for $w=1)$ the Taylor expansion of the nonlinearity function $F(w)$ yields $F(1)=0$ and $F'(1) \neq 0$. We note, however, that the theory remains valid for any real-valued function $F(w)$ that can be approximated  by a Taylor expansion in the vicinity of the EP. In this context, 
we consider separately four categories of functions
that yield elliptic umbilic EP structures
that include and extend  the example discussed in 
Sec. \ref{Sec:quadratic-nonlinearity} and \ref{Sec:extension-beyond-quadr-nonlin}. These are 
(1) $F(1) \neq 0$ and   $  F'(1) \neq 0$,
(2) $F(1)=0$     and   $F'(1) \neq 0$,
(3) $F(1) \neq 0$     and   $F'(1) = 0$, and
(4) $F(1) = 0$     and   $F'(1) = 0$.
Here category (2) is the one discussed above.
 We note that
 in the four categories
 the elliptic umbilic singularity sets emerge as functions of different parameters/variables. But common to these categories is the elliptic umbilic topology, in other words their singularity sets are topologically equivalent.
 Further details about the Lyapunov functions for the four categories are given in Sec I.B.1
of the Supplementary Information.
 Examples of matrices belonging to category (1) include the Hamiltonian in Eq. (1) of Ref. \cite{Bai-etal.2023} evaluated without noise, and Eq. (5) of Ref. \cite{zheng-chong.2025}.
}

\revisionC{\subsection{Asymmetric matrices, complex nonlinear coupling constants}\label{Sec:ecomplex-alpha}}

 \revisionC{We have also considered the case of asymmetric matrices, with the off-diagonal parameters $b$ and $c$ in the Hamiltonian Eq.
 \eqref{equ:Hnonlinear-cubic_1}
 being independent complex parameters. Our findings are summarized here, and the algebraic derivation is detailed in Sec. I.B.3 and I.B.4 in the Supplementary Information. As in the symmetric-matrix case, we construct a Lyapunov potential for the eigenvector elements of this asymmetric Hamiltonian in Sec. I.B.3. A straightforward expansion around the linear EP shows that the germ (the limiting Lyapunov potential as a function of ($r$, $\phi$) when the parameters are set at the EP values) has the structure of a double elliptic umbilic. We derive in Sec. I.B.4 a coordinate transformation that maps this structure into a canonical elliptic umbilic in the new coordinates. This shows that the (linear and nonlinear) EP sets also possess the topological structure of the elliptic umbilic around the EP.}

\revisionC{We find that the elliptic umbilic topology is also retained when the nonlinear coupling constant $\alpha$ is imaginary or complex. We comment briefly on this case in Sec. I.B.2 in the Supplementary Information.}

\revisionC{\subsection{Symmetry in nonlinear coefficient $\alpha$}\label{Sec:symmetry-in-alpha}}

\revisionC{We also mention again that the EP structure shown in Fig. \ref{fig:3D-elliptic-umbilic} is symmetric with respect to the 
Kerr-like nonlinearity $\alpha$ and the linear coupling $\beta$
(i.e., invariant under $\alpha \rightarrow - \alpha$, $\beta \rightarrow - \beta$).
For real-valued self ($\alpha_c$) and cross ($\alpha_x$) couplings in Eq. \eqref{equ:Hnonlinear-1}, it is instructive to compare our nonlinear $2 \times 2$ eigenvalue problem with the nonlinear Schr{\"{o}}dinger equation (NLS). If the NLS is derived from the Helmholtz equation in paraxial approximation, then the spatial propagation coordinate $z$ plays the role of time and the diffraction (second-order derivative in the transverse direction) plays the role of the kinetic energy, see e.g. \cite{saleh-teich.06}.
While the general solution structure of the NLS  includes effects like soliton formation and is generally complicated (and the NSL is not integrable in more than one spatial dimensions), it has been shown in  \cite{sterke-sipe.1990} that 
coupled-mode analysis in the nonlinear regime can be done with the NLS under suitable conditions. If one would use two modes with different transverse extensions, effects analogous to  self-focusing and self-defocusing could be described by a 2-mode model. In the language of our eigenvalue equation, the eigenvalue $E$ would correspond to the wave vector in the propagation direction $z$,
and the coupling coefficients $\alpha_c$ and $\alpha_x$ would depend on the spatial overlap between the two modes.
More detailed studies of the discrete NLS (also called discrete self-trapping equation) are provided in Ref. \cite{scott.2003}, but most studies of the discrete NLS, including 2-mode models that are equivalent to the (nonlinear) pendulum 
\cite{cruzeirohansson-etal.1988}, do not include decay, which is essential in our study of the EP structure.
If, however, 
a self-focusing and self-defocusing system with decay can be described by our 2-mode model, Eq. \eqref{equ:Hnonlinear-1}, with the difference in the absorption coefficients of the two modes given by  $\gamma$, then 
we predict that the EP structure is symmetric in $\alpha = \frac{1}{2} (\alpha_c - \alpha_x)$ as well as the linear coupling $\beta$
and the energy splitting $\delta$, as shown in Fig. \ref{fig:3D-elliptic-umbilic}.
}

\revisionC{\subsection{Larger matrix dimensions}\label{Sec:N-gt-2}}

For $N \times N$ matrices with $N > 2$, the dimension of the state variable space increases with $N$, the set of linear EPs is expected to have more structure than that of the $N=2$ case. More complicated, higher-order singularities, e.g. parabolic umbilics or singularities outside of the elementary catastrophe set, may be involved,
most of which have higher codimensions  (minimum dimension of the parameter space to capture the full structure of the singularity set) than the elliptic umbilic.
 But even when a relatively complete analysis comparable to the $2 \times 2$ case cannot be done, using the tools of catastrophe theory and  bifurcation theory may still provide fruitful insight into the topological/geometric structure of the singularity set.
\revisionC{We also note that in higher dimensional systems (described by $N \times N$ matrices with $N > 2$), even in the linear case we can have EPs of various orders, for example a $3 \times 3$ matrix can have second and third-order linear EPs. The  structure of the nonlinear EPs can then be expected to be more complex than the case of nonlinear EPs discussed above for the $2 \times 2$ matrix. We found that in the $2 \times 2$ case where a Lyapunov potential exists, the coalescence of eigenvectors agrees with the condition of singularities of the Lyapunov function, 
and - as is already known \cite{PhysRevApplied.18.054059,Bai-etal.2023,bai-etal.2024} -
the order of the nonlinear EPs can be higher than that of the linear EPs (in our case, the linear problem can only have second order EPs while the nonlinear problem has up to 4th-order nonlinear EPs). 
}

\revisionC{
\section{Discussion}
}

In summary, we have shown that the rich interplay of nonlinearity and non-Hermiticity at exceptional points 
\revisionC{results in a non-trivial topological structure of the EPs in the space of parameters including the nonlinearity parameter $\alpha$. This structure (or singularity set) is called elliptic umbilic and shown in Fig. \ref{fig:3D-elliptic-umbilic}.
Our findings apply to a large class of nonlinear systems that include -- but are not limited to --  all power-law nonlinearity functions $f(u)$ in the effective non-Hermitian Hamiltonian Eq.~\eqref{equ:H1nonlinear-cubic} with self-coupling and cross-coupling, where
the coupling coefficients can be real (in optics related to nonlinear refractive index coupling), imaginary (often related to nonlinear decay or gain), or complex. Asymmetric matrices with complex parameters in the off-diagonal elements also belong to this class.
Further extensions of the class of matrices with the elliptic umbilic topological structure may be possible. 
}

\revisionC{Specifically, we find that,
 by mapping EPs to degenerate critical points of a potential, catastrophe theory predicts that a second-order linear EP
 unfolds into the elliptic umbilic form. In practical terms, the EP perturbed by nonlinearity will spawn two touching cones of EPs with tri-cusped boundaries, independent of the underlying physical model. Interestingly, the non-trivial topological properties of the 
 set of EPs in the space of parameters including the nonlinearity parameter $\alpha$  can be predicted solely from the Lyapunov potential at the linear EP (i.e. with $\alpha=0$). This is a strong statement made possible by catastrophe theory, and in this sense, the topology can be regarded as `universal'.
 }
 From this topological viewpoint, the original linear EP acts as the organizing center of a higher-order singularity: a second-order EP becomes effectively fourth-order in the extended space of nonlinear parameters. These conclusions are the same for all physical platforms, e.g. photonic, atomic, and condensed-matter platforms.
Our approach can be extended to the case of \revisionC{third (or higher) order linear EPs  
(e.g. \cite{delplace-etal.2021,hu-etal.2023,lee-etal.2024,panahi-etal.2024})  } 
perturbed by nonlinearities.
\revisionC{Since, in the linear case, we find the germ to be that of an elliptic umbilic catastrophe, 
we conjecture that all  nonlinearities that are real or complex functions of the variables $w$ and $\theta$ defining the eigenvectors,
and that are analytic in a neighborhood around the linear EPs, will lead to locally elliptic umbilic singularity sets. 
}

Beyond the conceptual advance, our results have practical implications. Knowing the canonical cone--and--cusp shape of nonlinear EPs provides a clear blueprint for engineering non-Hermitian devices. For example, one can predict how tuning parameters (gain saturation, coupling strengths, etc.) will move or create EPs in polariton lasers, optical resonators, or PT-symmetric circuits. Crucially, because EP-based sensors and mode-control schemes depend on the topology of degeneracies, this universality 
will make it possible to establish rigorous bounds on the enhancement of sensitivity  of EPs in nonlinear systems
and enables
the design of robust, high-sensitivity devices across different platforms.
\revisionC{Our prediction of the singularity structure of EPs in the nonlinear regime should enable systematic studies and possible optimization of the renormalization of
EPs due to noise and stochastic processes, e.g. 
\cite{Bai-etal.2023,suntharalingam-etal.2023,panahi-etal.2024,zheng-chong.2025}, as well as studies of Liouvillian EPs
\cite{minganti-etal.2019,sun-yi.2024}.
}
From a broader perspective, identifying a common ``fingerprint'' for EPs highlights a deep unity in non-Hermitian physics. We expect these principles to guide future explorations -- for instance in higher-dimensional multi-mode or many-body setups -- and to inform the development of new spectroscopic and sensing techniques grounded in the universal topology of exceptional points. \\

%%%%%%%%%%%%    BEGIN  METHODS
\section{ Methods} 

The methodology of our theoretical approach is based on a branch of bifurcation theory called catastrophe theory. 
Here, we summarize the main aspects used in this method of analyzing nonlinear systems. More details are given in Sec. II of the Supplementary Information.

We begin with a brief review of catastrophe theory.
Consider a smooth function (Lyapunov potential)  $V_L \left( x_1, x_2, ... , x_n , a_1 , a_2 , ... , a_m \right)$ of  $n$  state variables $x_i, i = 1,...,n$ and $m$ control parameters $a_j,j=1,...,m$. The critical points of $V$ are points in state space where (solutions in state space to the equations)
\begin{equation}\label{critical-points-V.equ}
\frac {\partial V_L} {\partial x_i} = 0 \quad , \quad i = 1,...,n\,.
\end{equation}
Catastrophe theory seeks to classify the qualitative behavior of the critical points when the parameters are varied. If we think of the parameter space as a 'phase diagram', qualitative changes of critical point behavior happen at parameter space points for which $V$ has degenerate critical points (singularities). At these points (in parameter and state spaces), the Hessian matrix of second derivatives is singular,
\begin{equation}
{\rm det} \left[ \frac {\partial^2 V_L} {\partial x_i \partial x_j} \right] = 0\,,
\end{equation}
and multiple critical points coincide. We call the set of points in parameter space where $V_L$ has degenerate critical points the singularity set.
The theory classifies the structure of the singularity set locally around 'organizing points'.
Simple structures include folds, cusps, and umbilics.
Complete tables for the elementary catastrophes with codimension (defined below)
up to 5 were given in Refs. \cite{saunders.80,gilmore.81}. 
The table is reproduced as \revisionC{Supplementary Tab. I} in the Supplementary Information. 
For each catastrophe in the table, $x \equiv x_1$, $y \equiv x_2$ are state variables, and the other lower-case letters in the Perturbation column are parameters. For a given parameterized function $V_L$, an isolated point in the singularity set is fixed as the organizing point (in parameter space). At this parameter value, $V_L$ as a function of the state variables has one or more degenerate critical points, one of which is chosen for consideration. The state coordinates are redefined with the origin shifted to this degenerate critical point, and the value of $V_L$ is also shifted to zero at this point. To match a particular catastrophe, one may Taylor expand $V_L(x,y)$, with parameters set at the organizing point values, around the origin and compare the lowest-order terms with the standard forms in the Germ column in \revisionC{Supplementary Tab.\ I.} A smooth coordinate transformation may be needed to match the shown standard form. The singularity set, when restricted to the organizing point, may be structurally unstable: the geometric structure of the singularity set, and other properties, may change under small perturbations. The Perturbation column in \revisionC{Supplementary Tab.\ I.}  I shows some standard forms of perturbations that stabilize the singularity set: its geometric structure does not change under further small perturbations. The parameters equal zero at the organizing point. For each catastrophe, the number of parameters in the shown perturbation is the minimum number that can effect the stabilization. This minimum number is the codimension of the catastrophe.
A more mathematically precise explanation of Catastrophe theory terminology and concepts, such as germs 
and codimension, can be found in, for example, Ref. \cite{castrigiano-hayes.2004}.

Next, we discuss exceptional points of the (nonlinear) eigenvalue problem and the relation to catastrophe theory.
Consider an eigenvalue problem with an $N \times N$ matrix which is nonlinear in the eigenvector. Suppose the eigenvector is defined by $n$ real variables (a complex variable is considered as a pair of real variables), with $n \leq 2N$. Call these variables $x_i , i = 1,...,n$. Suppose also that the matrix depends on $m$ parameters $a_j, j=1,...,m$ and the eigenvector equation, after elimination of the eigenvalue, can be written as a set of $n$ equations
\begin{equation}\label{eigenvector.equ}
f_k \left( x_1,...,x_n,a_1,...,a_m \right) = 0 \quad , \quad k = 1,...,n
\end{equation}
For each fixed parameter set $\left( a_j , j=1,...,m \right)$, the solution $\left( x_i , i=1,...,n \right)$ to Eq.~(\ref{eigenvector.equ}) gives the eigenvector. The exceptional points are points in parameter space for which multiple eigenvectors, or equivalently, multiple solutions $\left( x_i , i=1,...,n \right)$ to Eq.~(\ref{eigenvector.equ}), coalesce.\\

\noindent
If a potential function $V_L \left( x_1, x_2, ... , x_n , a_1 , a_2 , ... , a_m \right)$ such that
\begin{equation}
f_k \left( x_1,...,x_n,a_1,...,a_m \right) = \frac {\partial V_L} {\partial x_k} \quad , \quad k = 1,...,n
\end{equation}
can be constructed, then the nonlinear eigenvalue problem can be 'mapped' onto the catastrophe theory formalism. The solutions $\left( x_i , i=1,...,n \right)$ to Eq. (\ref{eigenvector.equ}) are the critical points of $V_L$, and the machinery of catastrophe theory can be used to analyze the structure of the set of exceptional points, which is just the singularity set (parameter space subset where $V_L$ carries degenerate critical points). 
In our application, the linear EP is chosen as the organizing point (see previous subsection) and structurally stabilizing perturbations are identified.

\noindent
A necessary condition for the potential $V_L$ to exist is that the set $f_k$, $k = 1,...,n$ have 'vanishing curl': 
\begin{equation}\label{zero-curl.equ}
\frac {\partial f_i} {\partial x_j} - \frac {\partial f_j} {\partial x_i} = 0 \quad , \quad i,j = 1,...,n
\end{equation}
But for our purpose here, this condition can be slightly relaxed. Even if the $f_k$ in question does not satisfy Eq. (\ref{zero-curl.equ}), if a related set of functions, $f^\prime_k (x_1,...,x_n,a_1,...,a_m)$, $k=1,...,n$, has the same solutions when used in Eq. (\ref{eigenvector.equ}) and is the gradient of a potential $V_L$, then the critical-point analysis of $V_L$ is applicable to the EPs of $f_k$. This is the case for the eigenvector equations considered in this paper. The left hand sides of the real and imaginary parts of the eigenvector equations are equal to $w \partial V_L / \partial w$ and $- \partial V_L / \partial \theta$ respectively.

%%%%%%%%%%%%    END METHODS

\section{ Data availability}

\revisionC{
Source Data are provided with the paper. The main figures of this paper, Fig. \ref{fig:3D-elliptic-umbilic}, \ref{fig:elliptic-umbilic-cross-section} and \ref{fig:eigenvalues}, are available as data and png files
on GitHub via the DOI:
https://doi.org/10.5281/zenodo.17589377
}

\section{Code availability} 

\revisionC{
The Matlab code that generated the surface of the main figure (Fig. \ref{fig:3D-elliptic-umbilic}), cross sections of the surface (Fig. \ref{fig:elliptic-umbilic-cross-section}), and complex eigenvalue trajectories (Fig. \ref{fig:eigenvalues})
are available
on GitHub via the DOI:
https://doi.org/10.5281/zenodo.17589377
}

%\section{References} 

%%%%     REFERENCES
%\bibliographystyle{prsty}
%\bibliographystyle{prsty_noetal_noabbr}
%\bibliographystyle{osajnl-no-comma}

%this is redundant:
%\bibliographystyle{sn-mathphys-num}

%Rolf
%\bibliography{../../bib/allref}
%\bibliography{umbilic-references}

%\bibliography{allref}

\begin{thebibliography}{99}
\ifx \bisbn   \undefined \def \bisbn  #1{ISBN #1}\fi
\ifx \binits  \undefined \def \binits#1{#1}\fi
\ifx \bauthor  \undefined \def \bauthor#1{#1}\fi
\ifx \batitle  \undefined \def \batitle#1{#1}\fi
\ifx \bjtitle  \undefined \def \bjtitle#1{#1}\fi
\ifx \bvolume  \undefined \def \bvolume#1{\textbf{#1}}\fi
\ifx \byear  \undefined \def \byear#1{#1}\fi
\ifx \bissue  \undefined \def \bissue#1{#1}\fi
\ifx \bfpage  \undefined \def \bfpage#1{#1}\fi
\ifx \blpage  \undefined \def \blpage #1{#1}\fi
\ifx \burl  \undefined \def \burl#1{\textsf{#1}}\fi
\ifx \doiurl  \undefined \def \doiurl#1{\url{https://doi.org/#1}}\fi
\ifx \betal  \undefined \def \betal{\textit{et al.}}\fi
\ifx \binstitute  \undefined \def \binstitute#1{#1}\fi
\ifx \binstitutionaled  \undefined \def \binstitutionaled#1{#1}\fi
\ifx \bctitle  \undefined \def \bctitle#1{#1}\fi
\ifx \beditor  \undefined \def \beditor#1{#1}\fi
\ifx \bpublisher  \undefined \def \bpublisher#1{#1}\fi
\ifx \bbtitle  \undefined \def \bbtitle#1{#1}\fi
\ifx \bedition  \undefined \def \bedition#1{#1}\fi
\ifx \bseriesno  \undefined \def \bseriesno#1{#1}\fi
\ifx \blocation  \undefined \def \blocation#1{#1}\fi
\ifx \bsertitle  \undefined \def \bsertitle#1{#1}\fi
\ifx \bsnm \undefined \def \bsnm#1{#1}\fi
\ifx \bsuffix \undefined \def \bsuffix#1{#1}\fi
\ifx \bparticle \undefined \def \bparticle#1{#1}\fi
\ifx \barticle \undefined \def \barticle#1{#1}\fi
\bibcommenthead
\ifx \bconfdate \undefined \def \bconfdate #1{#1}\fi
\ifx \botherref \undefined \def \botherref #1{#1}\fi
\ifx \url \undefined \def \url#1{\textsf{#1}}\fi
\ifx \bchapter \undefined \def \bchapter#1{#1}\fi
\ifx \bbook \undefined \def \bbook#1{#1}\fi
\ifx \bcomment \undefined \def \bcomment#1{#1}\fi
\ifx \oauthor \undefined \def \oauthor#1{#1}\fi
\ifx \citeauthoryear \undefined \def \citeauthoryear#1{#1}\fi
\ifx \endbibitem  \undefined \def \endbibitem {}\fi
\ifx \bconflocation  \undefined \def \bconflocation#1{#1}\fi
\ifx \arxivurl  \undefined \def \arxivurl#1{\textsf{#1}}\fi
\csname PreBibitemsHook\endcsname

%%% 1
\bibitem[\protect\citeauthoryear{Kato}{1966}]{kato.1966}
\begin{bbook}
\bauthor{\bsnm{Kato}, \binits{T.}}:
\bbtitle{Perturbation Theory for Linear Operators}.
\bpublisher{Springer},
\blocation{New York, United States}
(\byear{1966})
\end{bbook}
\endbibitem

%%% 2
\bibitem[\protect\citeauthoryear{Seyranian et~al.}{2005}]{seyranian-etal.2005}
\begin{barticle}
\bauthor{\bsnm{Seyranian}, \binits{A.P.}},
\bauthor{\bsnm{Kirillov}, \binits{O.N.}},
\bauthor{\bsnm{Mailybaev}, \binits{A.A.}}:
\batitle{Coupling of eigenvalues of complex matrices at diabolic and exceptional points}.
\bjtitle{Journal of Physics A: Mathematical and General}
\bvolume{38},
\bfpage{1723}
(\byear{2005})
\end{barticle}
\endbibitem

%%% 3
\bibitem[\protect\citeauthoryear{Ashida et~al.}{2021}]{ashida-etal.2021}
\begin{barticle}
\bauthor{\bsnm{Ashida}, \binits{Y.}},
\bauthor{\bsnm{Gong}, \binits{Z.}},
\bauthor{\bsnm{Ueda}, \binits{M.}}:
\batitle{{Non-Hermitian physics}}.
\bjtitle{Advances in Physics}
\bvolume{69},
\bfpage{249}--\blpage{435}
(\byear{2021})
\end{barticle}
\endbibitem

%%% 4
\bibitem[\protect\citeauthoryear{Bender and Wu}{1969}]{bender-wu.1969}
\begin{barticle}
\bauthor{\bsnm{Bender}, \binits{C.M.}},
\bauthor{\bsnm{Wu}, \binits{T.T.}}:
\batitle{Anharmonic oscillator}.
\bjtitle{Phys. Rev.}
\bvolume{184},
\bfpage{1231}--\blpage{1260}
(\byear{1969})
\doiurl{10.1103/PhysRev.184.1231}
\end{barticle}
\endbibitem

%%% 5
\bibitem[\protect\citeauthoryear{Berry and O'Dell}{1998}]{berry-odell.1998}
\begin{barticle}
\bauthor{\bsnm{Berry}, \binits{M.V.}},
\bauthor{\bsnm{O'Dell}, \binits{D.H.J.}}:
\batitle{{Diffraction by volume gratings with imaginary potentials}}.
\bjtitle{{J. Phy. A: Math. Gen.}}
\bvolume{31},
\bfpage{2093}
(\byear{1998})
\end{barticle}
\endbibitem

%%% 6
\bibitem[\protect\citeauthoryear{Leyvraz and Heiss}{2005}]{leyvraz-heiss.2005}
\begin{barticle}
\bauthor{\bsnm{Leyvraz}, \binits{F.}},
\bauthor{\bsnm{Heiss}, \binits{W.D.}}:
\batitle{Large-$n$ scaling behavior of the lipkin-meshkov-glick model}.
\bjtitle{Phys. Rev. Lett.}
\bvolume{95},
\bfpage{050402}
(\byear{2005})
\doiurl{10.1103/PhysRevLett.95.050402}
\end{barticle}
\endbibitem

%%% 7
\bibitem[\protect\citeauthoryear{Heiss}{2012}]{heiss.2012}
\begin{barticle}
\bauthor{\bsnm{Heiss}, \binits{W.D.}}:
\batitle{{The physics of exceptional points}}.
\bjtitle{Journal of Physics A: Mathematical and Theoretical}
\bvolume{45},
\bfpage{444016}
(\byear{2012})
\end{barticle}
\endbibitem

%%% 8
\bibitem[\protect\citeauthoryear{Brandstetter et~al.}{2014}]{brandstetter-etal.2014}
\begin{barticle}
\bauthor{\bsnm{Brandstetter}, \binits{M.}},
\bauthor{\bsnm{Liertzer}, \binits{M.}},
\bauthor{\bsnm{Deutsch}, \binits{C.}},
\bauthor{\bsnm{Klang}, \binits{P.}},
\bauthor{\bsnm{Schoberl}, \binits{J.}},
\bauthor{\bsnm{Tureci}, \binits{H.E.}},
\bauthor{\bsnm{Strasser}, \binits{G.}},
\bauthor{\bsnm{Unterrainer}, \binits{K.}},
\bauthor{\bsnm{Rotter}, \binits{S.}}:
\batitle{{Reversing the pump dependence of a laser at an exceptional point}}.
\bjtitle{Nat. Commun.}
\bvolume{5},
\bfpage{4034}
(\byear{2014})
\end{barticle}
\endbibitem

%%% 9
\bibitem[\protect\citeauthoryear{Liertzer et~al.}{2012}]{lietzer-etal.2012}
\begin{barticle}
\bauthor{\bsnm{Liertzer}, \binits{M.}},
\bauthor{\bsnm{Ge}, \binits{L.}},
\bauthor{\bsnm{Cerjan}, \binits{A.}},
\bauthor{\bsnm{Stone}, \binits{A.D.}},
\bauthor{\bsnm{T\"ureci}, \binits{H.E.}},
\bauthor{\bsnm{Rotter}, \binits{S.}}:
\batitle{Pump-induced exceptional points in lasers}.
\bjtitle{Phys. Rev. Lett.}
\bvolume{108},
\bfpage{173901}
(\byear{2012})
\doiurl{10.1103/PhysRevLett.108.173901}
\end{barticle}
\endbibitem

%%% 10
\bibitem[\protect\citeauthoryear{Kullig et~al.}{2023}]{kullig-etal.2023}
\begin{barticle}
\bauthor{\bsnm{Kullig}, \binits{J.}},
\bauthor{\bsnm{Grom}, \binits{D.}},
\bauthor{\bsnm{Klembt}, \binits{S.}},
\bauthor{\bsnm{Wiersig}, \binits{J.}}:
\batitle{Higher-order exceptional points in waveguide-coupled microcavities: perturbation induced frequency splitting and mode patterns}.
\bjtitle{Photon. Res.}
\bvolume{11},
\bfpage{54}--\blpage{64}
(\byear{2023})
\doiurl{10.1364/PRJ.496414}
\end{barticle}
\endbibitem

%%% 11
\bibitem[\protect\citeauthoryear{Longhi}{2018}]{longhi.2018}
\begin{barticle}
\bauthor{\bsnm{Longhi}, \binits{S.}}:
\batitle{Exceptional points and photonic catastrophe}.
\bjtitle{Opt. Lett.}
\bvolume{43},
\bfpage{2929}--\blpage{2932}
(\byear{2018})
\doiurl{10.1364/OL.43.002929}
\end{barticle}
\endbibitem

%%% 12
\bibitem[\protect\citeauthoryear{Quiroz-Ju\'{a}rez et~al.}{2019}]{quirozjuarez.2019}
\begin{barticle}
\bauthor{\bsnm{Quiroz-Ju\'{a}rez}, \binits{M.A.}},
\bauthor{\bsnm{Perez-Leija}, \binits{A.}},
\bauthor{\bsnm{Tschernig}, \binits{K.}},
\bauthor{\bsnm{Rodr\'{i}guez-Lara}, \binits{B.M.}},
\bauthor{\bsnm{{n}a-Loaiza}, \binits{O.S.M.}},
\bauthor{\bsnm{Busch}, \binits{K.}},
\bauthor{\bsnm{Joglekar}, \binits{Y.N.}},
\bauthor{\bsnm{J.~Le\'{o}n-Montiel}, \binits{R.}}:
\batitle{Exceptional points of any order in a single, lossy waveguide beam splitter by photon-number-resolved detection}.
\bjtitle{Photon. Res.}
\bvolume{7},
\bfpage{862}--\blpage{867}
(\byear{2019})
\doiurl{10.1364/PRJ.7.000862}
\end{barticle}
\endbibitem

%%% 13
\bibitem[\protect\citeauthoryear{Dong et~al.}{2020}]{dong-etal.2020}
\begin{barticle}
\bauthor{\bsnm{Dong}, \binits{S.}},
\bauthor{\bsnm{Hu}, \binits{G.}},
\bauthor{\bsnm{Wang}, \binits{Q.}},
\bauthor{\bsnm{Jia}, \binits{Y.}},
\bauthor{\bsnm{Zhang}, \binits{Q.}},
\bauthor{\bsnm{Cao}, \binits{G.}},
\bauthor{\bsnm{Wang}, \binits{J.}},
\bauthor{\bsnm{Chen}, \binits{S.}},
\bauthor{\bsnm{Fan}, \binits{D.}},
\bauthor{\bsnm{Jiang}, \binits{W.}}, \betal:
\batitle{Loss-assisted metasurface at an exceptional point}.
\bjtitle{ACS Photonics}
\bvolume{7},
\bfpage{3321}--\blpage{3327}
(\byear{2020})
\end{barticle}
\endbibitem

%%% 14
\bibitem[\protect\citeauthoryear{Xu et~al.}{2017}]{xu-etal.2017}
\begin{barticle}
\bauthor{\bsnm{Xu}, \binits{Y.}},
\bauthor{\bsnm{Wang}, \binits{S.-T.}},
\bauthor{\bsnm{Duan}, \binits{L.-M.}}:
\batitle{Weyl exceptional rings in a three-dimensional dissipative cold atomic gas}.
\bjtitle{Phys. Rev. Lett.}
\bvolume{118},
\bfpage{045701}
(\byear{2017})
\doiurl{10.1103/PhysRevLett.118.045701}
\end{barticle}
\endbibitem

%%% 15
\bibitem[\protect\citeauthoryear{Choi et~al.}{2010}]{choi-etal.10}
\begin{barticle}
\bauthor{\bsnm{Choi}, \binits{Y.}},
\bauthor{\bsnm{Kang}, \binits{S.}},
\bauthor{\bsnm{Lim}, \binits{S.}},
\bauthor{\bsnm{Kim}, \binits{W.}},
\bauthor{\bsnm{Kim}, \binits{J.-R.}},
\bauthor{\bsnm{Lee}, \binits{J.-H.}},
\bauthor{\bsnm{An}, \binits{K.}}:
\batitle{{Quasieigenstate coalescence in an atom-cavity quantum composite}}.
\bjtitle{Phys. Rev. Lett.}
\bvolume{104},
\bfpage{153601}
(\byear{2010})
\end{barticle}
\endbibitem

%%% 16
\bibitem[\protect\citeauthoryear{Wang et~al.}{2024}]{wang-etal.2024}
\begin{barticle}
\bauthor{\bsnm{Wang}, \binits{C.}},
\bauthor{\bsnm{Li}, \binits{N.}},
\bauthor{\bsnm{Xie}, \binits{J.}},
\bauthor{\bsnm{Ding}, \binits{C.}},
\bauthor{\bsnm{Ji}, \binits{Z.}},
\bauthor{\bsnm{Xiao}, \binits{L.}},
\bauthor{\bsnm{Jia}, \binits{S.}},
\bauthor{\bsnm{Yan}, \binits{B.}},
\bauthor{\bsnm{Hu}, \binits{Y.}},
\bauthor{\bsnm{Zhao}, \binits{Y.}}:
\batitle{Exceptional nexus in bose-einstein condensates with collective dissipation}.
\bjtitle{Phys. Rev. Lett.}
\bvolume{132},
\bfpage{253401}
(\byear{2024})
\doiurl{10.1103/PhysRevLett.132.253401}
\end{barticle}
\endbibitem

%%% 17
\bibitem[\protect\citeauthoryear{Dembowski et~al.}{2001}]{dembowski-etal.01}
\begin{barticle}
\bauthor{\bsnm{Dembowski}, \binits{C.}},
\bauthor{\bsnm{Graf}, \binits{H.-D.}},
\bauthor{\bsnm{Harney}, \binits{H.L.}},
\bauthor{\bsnm{Heine}, \binits{A.}},
\bauthor{\bsnm{Heiss}, \binits{W.D.}},
\bauthor{\bsnm{Rehfeld}, \binits{H.}},
\bauthor{\bsnm{Richter}, \binits{A.}}:
\batitle{{Experimental observation of the topological structure of exceptional points}}.
\bjtitle{Phys. Rev. Lett.}
\bvolume{86},
\bfpage{787}--\blpage{790}
(\byear{2001})
\end{barticle}
\endbibitem

%%% 18
\bibitem[\protect\citeauthoryear{Miao et~al.}{2016}]{miao-etal.2016}
\begin{barticle}
\bauthor{\bsnm{Miao}, \binits{P.}},
\bauthor{\bsnm{Zhang}, \binits{Z.}},
\bauthor{\bsnm{Sun}, \binits{J.}},
\bauthor{\bsnm{Walasik}, \binits{W.}},
\bauthor{\bsnm{Longhi}, \binits{S.}},
\bauthor{\bsnm{Litchinitser}, \binits{N.M.}},
\bauthor{\bsnm{Feng}, \binits{L.}}:
\batitle{{Orbital angular momentum microlaser}}.
\bjtitle{Science}
\bvolume{535},
\bfpage{464}
(\byear{2016})
\end{barticle}
\endbibitem

%%% 19
\bibitem[\protect\citeauthoryear{Gao et~al.}{2015}]{gao-etal.15}
\begin{barticle}
\bauthor{\bsnm{Gao}, \binits{T.}},
\bauthor{\bsnm{Estrecho}, \binits{E.}},
\bauthor{\bsnm{Bliokh}, \binits{K.Y.}},
\bauthor{\bsnm{Liew}, \binits{T.C.H.}},
\bauthor{\bsnm{Fraser}, \binits{M.D.}},
\bauthor{\bsnm{Brodbeck}, \binits{S.}},
\bauthor{\bsnm{Kamp}, \binits{M.}},
\bauthor{\bsnm{Schneider}, \binits{C.}},
\bauthor{\bsnm{Hoefling}, \binits{S.}},
\bauthor{\bsnm{Yamamoto}, \binits{Y.}},
\bauthor{\bsnm{Nori}, \binits{F.}},
\bauthor{\bsnm{Kivshar}, \binits{Y.S.}},
\bauthor{\bsnm{Truscott}, \binits{A.G.}},
\bauthor{\bsnm{Dall}, \binits{R.G.}},
\bauthor{\bsnm{Ostrovskay}, \binits{E.A.}}:
\batitle{{Observation of non-Hermitian degeneracies in a chaotic exciton-polariton billiard}}.
\bjtitle{Nature}
\bvolume{526},
\bfpage{554}--\blpage{558}
(\byear{2015})
\end{barticle}
\endbibitem

%%% 20
\bibitem[\protect\citeauthoryear{Gao et~al.}{2018}]{gao2018chiral}
\begin{barticle}
\bauthor{\bsnm{Gao}, \binits{T.}},
\bauthor{\bsnm{Li}, \binits{G.}},
\bauthor{\bsnm{Estrecho}, \binits{E.}},
\bauthor{\bsnm{Liew}, \binits{T.C.H.}},
\bauthor{\bsnm{Comber-Todd}, \binits{D.}},
\bauthor{\bsnm{Nalitov}, \binits{A.}},
\bauthor{\bsnm{Steger}, \binits{M.}},
\bauthor{\bsnm{West}, \binits{K.}},
\bauthor{\bsnm{Pfeiffer}, \binits{L.}},
\bauthor{\bsnm{Snoke}, \binits{D.W.}},
\bauthor{\bsnm{Kavokin}, \binits{A.V.}},
\bauthor{\bsnm{Truscott}, \binits{A.G.}},
\bauthor{\bsnm{Ostrovskaya}, \binits{E.A.}}:
\batitle{Chiral modes at exceptional points in exciton-polariton quantum fluids}.
\bjtitle{Phys. {R}ev. {L}ett.}
\bvolume{120},
\bfpage{065301}
(\byear{2018})
\end{barticle}
\endbibitem

%%% 21
\bibitem[\protect\citeauthoryear{Ozturk et~al.}{2021}]{ozturk-etal.2021}
\begin{barticle}
\bauthor{\bsnm{Ozturk}, \binits{F.E.}},
\bauthor{\bsnm{Lappe}, \binits{T.}},
\bauthor{\bsnm{Hellmann}, \binits{G.}},
\bauthor{\bsnm{Schmitt}, \binits{J.}},
\bauthor{\bsnm{Klaers}, \binits{J.}},
\bauthor{\bsnm{Vewinger}, \binits{F.}},
\bauthor{\bsnm{Kroha}, \binits{J.}},
\bauthor{\bsnm{Weitz}, \binits{M.}}:
\batitle{{Observation of a non-Hermitian phase transition in an optical quantum gas}}.
\bjtitle{Science}
\bvolume{372},
\bfpage{88}--\blpage{91}
(\byear{2021})
\end{barticle}
\endbibitem

%%% 22
\bibitem[\protect\citeauthoryear{Khurgin}{2020}]{khurgin.2020}
\begin{barticle}
\bauthor{\bsnm{Khurgin}, \binits{J.}}:
\batitle{{Exceptional points in polaritonic cavities and subthreshold Fabry Perot lasers}}.
\bjtitle{Optica}
\bvolume{7},
\bfpage{1015}--\blpage{1023}
(\byear{2020})
\end{barticle}
\endbibitem

%%% 23
\bibitem[\protect\citeauthoryear{Li et~al.}{2022}]{Li2022}
\begin{barticle}
\bauthor{\bsnm{Li}, \binits{Y.}},
\bauthor{\bsnm{Ma}, \binits{X.}},
\bauthor{\bsnm{Hatzopoulos}, \binits{Z.}},
\bauthor{\bsnm{Savvidis}, \binits{P.G.}},
\bauthor{\bsnm{Schumacher}, \binits{S.}},
\bauthor{\bsnm{Gao}, \binits{T.}}:
\batitle{Switching off a microcavity polariton condensate near the exceptional point}.
\bjtitle{ACS Photonics}
\bvolume{9},
\bfpage{2079}--\blpage{2086}
(\byear{2022})
\doiurl{10.1021/acsphotonics.2c00288}
\end{barticle}
\endbibitem

%%% 24
\bibitem[\protect\citeauthoryear{Hanai et~al.}{2019}]{PhysRevLett.122.185301}
\begin{barticle}
\bauthor{\bsnm{Hanai}, \binits{R.}},
\bauthor{\bsnm{Edelman}, \binits{A.}},
\bauthor{\bsnm{Ohashi}, \binits{Y.}},
\bauthor{\bsnm{Littlewood}, \binits{P.B.}}:
\batitle{Non-{H}ermitian phase transition from a polariton {B}ose-{E}instein condensate to a photon laser}.
\bjtitle{Phys. Rev. Lett.}
\bvolume{122},
\bfpage{185301}
(\byear{2019})
\doiurl{10.1103/PhysRevLett.122.185301}
\end{barticle}
\endbibitem

%%% 25
\bibitem[\protect\citeauthoryear{Yu et~al.}{2021}]{PhysRevB.104.235408}
\begin{barticle}
\bauthor{\bsnm{Yu}, \binits{Z.-F.}},
\bauthor{\bsnm{Xue}, \binits{J.-K.}},
\bauthor{\bsnm{Zhuang}, \binits{L.}},
\bauthor{\bsnm{Zhao}, \binits{J.}},
\bauthor{\bsnm{Liu}, \binits{W.-M.}}:
\batitle{Non-{H}ermitian spectrum and multistability in exciton-polariton condensates}.
\bjtitle{Phys. Rev. B}
\bvolume{104},
\bfpage{235408}
(\byear{2021})
\doiurl{10.1103/PhysRevB.104.235408}
\end{barticle}
\endbibitem

%%% 26
\bibitem[\protect\citeauthoryear{Graefe et~al.}{2008}]{graefe-etal.2008}
\begin{barticle}
\bauthor{\bsnm{Graefe}, \binits{E.M.}},
\bauthor{\bsnm{Gunther}, \binits{U.}},
\bauthor{\bsnm{Korsch}, \binits{H.J.}},
\bauthor{\bsnm{Niederle}, \binits{A.E.}}:
\batitle{{A non-Hermitian symmetric Bose-Hubbard model: eigenvalue rings from unfolding higher-order exceptional points}}.
\bjtitle{Journal of Physics A: Mathematical and Theoretical.}
\bvolume{41},
\bfpage{255206}
(\byear{2008})
\end{barticle}
\endbibitem

%%% 27
\bibitem[\protect\citeauthoryear{Hanai and Littlewood}{2020}]{hanai-littlewood.20}
\begin{barticle}
\bauthor{\bsnm{Hanai}, \binits{R.}},
\bauthor{\bsnm{Littlewood}, \binits{P.}}:
\batitle{{Critical fluctuations at a many-body exceptional point}}.
\bjtitle{Phys. Rev. Research}
\bvolume{2},
\bfpage{033018}
(\byear{2020})
\end{barticle}
\endbibitem

%%% 28
\bibitem[\protect\citeauthoryear{Binder and Kwong}{2021}]{binder-kwong.2021}
\begin{barticle}
\bauthor{\bsnm{Binder}, \binits{R.}},
\bauthor{\bsnm{Kwong}, \binits{N.H.}}:
\batitle{Metamorphosis of goldstone and soft fluctuation modes in polariton lasers}.
\bjtitle{Phys. Rev. B}
\bvolume{103},
\bfpage{085304}
(\byear{2021})
\doiurl{10.1103/PhysRevB.103.085304}
\end{barticle}
\endbibitem

%%% 29
\bibitem[\protect\citeauthoryear{Wiersig}{2014}]{wiersig.2014}
\begin{barticle}
\bauthor{\bsnm{Wiersig}, \binits{J.}}:
\batitle{Enhancing the sensitivity of frequency and energy splitting detection by using exceptional points: application to microcavity sensors for single-particle detection}.
\bjtitle{Phys. Rev. Lett.}
\bvolume{112},
\bfpage{203901}
(\byear{2014})
\doiurl{10.1103/PhysRevLett.112.203901}
\end{barticle}
\endbibitem

%%% 30
\bibitem[\protect\citeauthoryear{Hodaei et~al.}{2017}]{hodaei-etal.2017}
\begin{barticle}
\bauthor{\bsnm{Hodaei}, \binits{H.}},
\bauthor{\bsnm{Hassan}, \binits{A.U.}},
\bauthor{\bsnm{Wittek}, \binits{S.}},
\bauthor{\bsnm{Garcia-Gracia}, \binits{H.}},
\bauthor{\bsnm{El-Ganainy}, \binits{R.}},
\bauthor{\bsnm{Christodoulides}, \binits{D.N.}},
\bauthor{\bsnm{Khajavikhan}, \binits{M.}}:
\batitle{{Enhanced sensitivity at higher-order exceptional points}}.
\bjtitle{Nature}
\bvolume{548},
\bfpage{187}
(\byear{2017})
\end{barticle}
\endbibitem

%%% 31
\bibitem[\protect\citeauthoryear{Langbein}{2018}]{langbein.2018}
\begin{barticle}
\bauthor{\bsnm{Langbein}, \binits{W.}}:
\batitle{No exceptional precision of exceptional-point sensors}.
\bjtitle{Phys. Rev. A}
\bvolume{98},
\bfpage{023805}
(\byear{2018})
\doiurl{10.1103/PhysRevA.98.023805}
\end{barticle}
\endbibitem

%%% 32
\bibitem[\protect\citeauthoryear{Wiersig}{2020}]{wiersig.2020}
\begin{barticle}
\bauthor{\bsnm{Wiersig}, \binits{J.}}:
\batitle{Review of exceptional point-based sensors}.
\bjtitle{Photon. Res.}
\bvolume{8},
\bfpage{1457}--\blpage{1467}
(\byear{2020})
\doiurl{10.1364/PRJ.396115}
\end{barticle}
\endbibitem

%%% 33
\bibitem[\protect\citeauthoryear{Sakhdari et~al.}{2019}]{sakhdari-etal.2019}
\begin{barticle}
\bauthor{\bsnm{Sakhdari}, \binits{M.}},
\bauthor{\bsnm{Hajizadegan}, \binits{M.}},
\bauthor{\bsnm{Zhong}, \binits{Q.}},
\bauthor{\bsnm{Christodoulides}, \binits{D.N.}},
\bauthor{\bsnm{El-Ganainy}, \binits{R.}},
\bauthor{\bsnm{Chen}, \binits{P.-Y.}}:
\batitle{Experimental observation of $pt$ symmetry breaking near divergent exceptional points}.
\bjtitle{Phys. Rev. Lett.}
\bvolume{123},
\bfpage{193901}
(\byear{2019})
\doiurl{10.1103/PhysRevLett.123.193901}
\end{barticle}
\endbibitem

%%% 34
\bibitem[\protect\citeauthoryear{Zheng and Chong}{2025}]{zheng-chong.2025}
\begin{barticle}
\bauthor{\bsnm{Zheng}, \binits{X.}},
\bauthor{\bsnm{Chong}, \binits{Y.D.}}:
\batitle{Noise constraints for nonlinear exceptional point sensing}.
\bjtitle{Phys. Rev. Lett.}
\bvolume{134},
\bfpage{133801}
(\byear{2025})
\doiurl{10.1103/PhysRevLett.134.133801}
\end{barticle}
\endbibitem

%%% 35
\bibitem[\protect\citeauthoryear{Bender and Hook}{2024}]{bender-hook.2024}
\begin{barticle}
\bauthor{\bsnm{Bender}, \binits{C.M.}},
\bauthor{\bsnm{Hook}, \binits{D.W.}}:
\batitle{$\mathcal{PT}$-symmetric quantum mechanics}.
\bjtitle{Rev. Mod. Phys.}
\bvolume{96},
\bfpage{045002}
(\byear{2024})
\doiurl{10.1103/RevModPhys.96.045002}
\end{barticle}
\endbibitem

%%% 36
\bibitem[\protect\citeauthoryear{Hassan et~al.}{2015}]{hassan-etal.2015}
\begin{barticle}
\bauthor{\bsnm{Hassan}, \binits{A.U.}},
\bauthor{\bsnm{Hodaei}, \binits{H.}},
\bauthor{\bsnm{Miri}, \binits{M.-A.}},
\bauthor{\bsnm{Khajavikhan}, \binits{M.}},
\bauthor{\bsnm{Christodoulides}, \binits{D.N.}}:
\batitle{Nonlinear reversal of the $\mathcal{PT}$-symmetric phase transition in a system of coupled semiconductor microring resonators}.
\bjtitle{Phys. Rev. A}
\bvolume{92},
\bfpage{063807}
(\byear{2015})
\doiurl{10.1103/PhysRevA.92.063807}
\end{barticle}
\endbibitem

%%% 37
\bibitem[\protect\citeauthoryear{Miri and Alu}{2016}]{miri-alu.2016}
\begin{barticle}
\bauthor{\bsnm{Miri}, \binits{M.-A.}},
\bauthor{\bsnm{Alu}, \binits{A.}}:
\batitle{Nonlinearity-induced pt-symmetry without material gain}.
\bjtitle{New Journal of Physics}
\bvolume{18},
\bfpage{065001}
(\byear{2016})
\doiurl{10.1088/1367-2630/18/6/065001}
\end{barticle}
\endbibitem

%%% 38
\bibitem[\protect\citeauthoryear{Ge and El-Ganainy}{2016}]{ge-elganainy.2016}
\begin{barticle}
\bauthor{\bsnm{Ge}, \binits{L.}},
\bauthor{\bsnm{El-Ganainy}, \binits{R.}}:
\batitle{Nonlinear modal interactions in parity-time (pt) symmetric lasers}.
\bjtitle{Scientific Reports}
\bvolume{6},
\bfpage{24889}
(\byear{2016})
\end{barticle}
\endbibitem

%%% 39
\bibitem[\protect\citeauthoryear{Kominis et~al.}{2017}]{kominis-etal.2017}
\begin{barticle}
\bauthor{\bsnm{Kominis}, \binits{Y.}},
\bauthor{\bsnm{Kovanis}, \binits{V.}},
\bauthor{\bsnm{Bountis}, \binits{T.}}:
\batitle{Controllable asymmetric phase-locked states of the fundamental active photonic dimer}.
\bjtitle{Phys. Rev. A}
\bvolume{96},
\bfpage{043836}
(\byear{2017})
\doiurl{10.1103/PhysRevA.96.043836}
\end{barticle}
\endbibitem

%%% 40
\bibitem[\protect\citeauthoryear{Teimourpour et~al.}{2017}]{teimourpour-etal.2017}
\begin{barticle}
\bauthor{\bsnm{Teimourpour}, \binits{M.}},
\bauthor{\bsnm{Khajavikhan}, \binits{M.}},
\bauthor{\bsnm{Christodoulides}, \binits{D.N.}},
\bauthor{\bsnm{El-Ganainy}, \binits{R.}}:
\batitle{Robustness and mode selectivity in parity-time (pt) symmetric lasers}.
\bjtitle{Scientific Reports}
\bvolume{7},
\bfpage{10756}
(\byear{2017})
\end{barticle}
\endbibitem

%%% 41
\bibitem[\protect\citeauthoryear{Ju et~al.}{2019}]{ju-etal.2019}
\begin{barticle}
\bauthor{\bsnm{Ju}, \binits{C.-Y.}},
\bauthor{\bsnm{Miranowicz}, \binits{A.}},
\bauthor{\bsnm{Chen}, \binits{G.-Y.}},
\bauthor{\bsnm{Nori}, \binits{F.}}:
\batitle{Non-hermitian hamiltonians and no-go theorems in quantum information}.
\bjtitle{Phys. Rev. A}
\bvolume{100},
\bfpage{062118}
(\byear{2019})
\end{barticle}
\endbibitem

%%% 42
\bibitem[\protect\citeauthoryear{Xia et~al.}{2021}]{xia-etal.2021}
\begin{barticle}
\bauthor{\bsnm{Xia}, \binits{S.}},
\bauthor{\bsnm{Kaltsas}, \binits{D.}},
\bauthor{\bsnm{Song}, \binits{D.}},
\bauthor{\bsnm{Komis}, \binits{I.}},
\bauthor{\bsnm{Xu}, \binits{J.}},
\bauthor{\bsnm{Szameit}, \binits{A.}},
\bauthor{\bsnm{Buljan}, \binits{H.}},
\bauthor{\bsnm{Makris}, \binits{K.G.}},
\bauthor{\bsnm{Chen}, \binits{Z.}}:
\batitle{Nonlinear tuning of pt symmetry and non-hermitian topological states}.
\bjtitle{Science}
\bvolume{372},
\bfpage{72}--\blpage{76}
(\byear{2021})
\doiurl{10.1126/science.abf6873}
\end{barticle}
\endbibitem

%%% 43
\bibitem[\protect\citeauthoryear{Lee et~al.}{2024}]{lee-etal.2024}
\begin{barticle}
\bauthor{\bsnm{Lee}, \binits{C.}},
\bauthor{\bsnm{Zhang}, \binits{K.}},
\bauthor{\bsnm{Miao}, \binits{J.}},
\bauthor{\bsnm{Sun}, \binits{K.}},
\bauthor{\bsnm{Deng}, \binits{H.}}:
\batitle{Topologically protected exceptional points and reentrant $\mathcal{PT}$ phase in an exact ternary model}.
\bjtitle{Phys. Rev. A}
\bvolume{109},
\bfpage{053503}
(\byear{2024})
\doiurl{10.1103/PhysRevA.109.053503}
\end{barticle}
\endbibitem

%%% 44
\bibitem[\protect\citeauthoryear{Klauck et~al.}{2025}]{klauck-etal.2025}
\begin{barticle}
\bauthor{\bsnm{Klauck}, \binits{F.U.J.}},
\bauthor{\bsnm{Heinrich}, \binits{M.}},
\bauthor{\bsnm{Szameit}, \binits{A.}},
\bauthor{\bsnm{Wolterink}, \binits{T.A.W.}}:
\batitle{Crossing exceptional points in non-hermitian quantum systems}.
\bjtitle{Science Advances}
\bvolume{11},
\bfpage{8275}
(\byear{2025})
\doiurl{10.1126/sciadv.adr8275}
\end{barticle}
\endbibitem

%%% 45
\bibitem[\protect\citeauthoryear{Abbasi et~al.}{2022}]{abbasi-etal.2022}
\begin{barticle}
\bauthor{\bsnm{Abbasi}, \binits{M.}},
\bauthor{\bsnm{Chen}, \binits{W.}},
\bauthor{\bsnm{Naghiloo}, \binits{M.}},
\bauthor{\bsnm{Joglekar}, \binits{Y.N.}},
\bauthor{\bsnm{Murch}, \binits{K.W.}}:
\batitle{Topological quantum state control through exceptional-point proximity}.
\bjtitle{Phys. Rev. Lett.}
\bvolume{128},
\bfpage{160401}
(\byear{2022})
\doiurl{10.1103/PhysRevLett.128.160401}
\end{barticle}
\endbibitem

%%% 46
\bibitem[\protect\citeauthoryear{Blinova et~al.}{2024}]{blinova-etal.2024}
\begin{barticle}
\bauthor{\bsnm{Blinova}, \binits{P.}},
\bauthor{\bsnm{Moiseev}, \binits{E.}},
\bauthor{\bsnm{Wang}, \binits{K.}}:
\batitle{Exceptional swallowtail degeneracies in driven-dissipative quadrature squeezing}.
\bjtitle{Phys. Rev. Res.}
\bvolume{6},
\bfpage{043209}
(\byear{2024})
\doiurl{10.1103/PhysRevResearch.6.043209}
\end{barticle}
\endbibitem

%%% 47
\bibitem[\protect\citeauthoryear{Luitz and Piazza}{2019}]{luitz-etal.2019}
\begin{barticle}
\bauthor{\bsnm{Luitz}, \binits{D.J.}},
\bauthor{\bsnm{Piazza}, \binits{F.}}:
\batitle{Exceptional points and the topology of quantum many-body spectra}.
\bjtitle{Phys. Rev. Res.}
\bvolume{1},
\bfpage{033051}
(\byear{2019})
\doiurl{10.1103/PhysRevResearch.1.033051}
\end{barticle}
\endbibitem

%%% 48
\bibitem[\protect\citeauthoryear{Kawabata et~al.}{2019}]{kawabata-etal.2019}
\begin{barticle}
\bauthor{\bsnm{Kawabata}, \binits{K.}},
\bauthor{\bsnm{Bessho}, \binits{T.}},
\bauthor{\bsnm{Sato}, \binits{M.}}:
\batitle{Classification of exceptional points and non-hermitian topological semimetals}.
\bjtitle{Phys. Rev. Lett.}
\bvolume{123},
\bfpage{066405}
(\byear{2019})
\doiurl{10.1103/PhysRevLett.123.066405}
\end{barticle}
\endbibitem

%%% 49
\bibitem[\protect\citeauthoryear{Xiao et~al.}{2020}]{xiao-etal.2020}
\begin{barticle}
\bauthor{\bsnm{Xiao}, \binits{Y.-X.}},
\bauthor{\bsnm{Ding}, \binits{K.}},
\bauthor{\bsnm{Zhang}, \binits{R.-Y.}},
\bauthor{\bsnm{Hang}, \binits{Z.H.}},
\bauthor{\bsnm{Chan}, \binits{C.T.}}:
\batitle{Exceptional points make an astroid in non-hermitian lieb lattice: Evolution and topological protection}.
\bjtitle{Phys. Rev. B}
\bvolume{102},
\bfpage{245144}
(\byear{2020})
\doiurl{10.1103/PhysRevB.102.245144}
\end{barticle}
\endbibitem

%%% 50
\bibitem[\protect\citeauthoryear{Tang et~al.}{2020}]{tang-etal.2020}
\begin{barticle}
\bauthor{\bsnm{Tang}, \binits{W.}},
\bauthor{\bsnm{Jiang}, \binits{X.}},
\bauthor{\bsnm{Ding}, \binits{K.}},
\bauthor{\bsnm{Xiao}, \binits{Y.}},
\bauthor{\bsnm{Zhang}, \binits{Z.}},
\bauthor{\bsnm{Chan}, \binits{C.}},
\bauthor{\bsnm{Ma}, \binits{G.}}:
\batitle{Exceptional nexus with a hybrid topological invariant}.
\bjtitle{Science}
\bvolume{370},
\bfpage{1077}--\blpage{1080}
(\byear{2020})
\end{barticle}
\endbibitem

%%% 51
\bibitem[\protect\citeauthoryear{Bergholtz et~al.}{2021}]{bergholtz-etal.2021}
\begin{barticle}
\bauthor{\bsnm{Bergholtz}, \binits{E.J.}},
\bauthor{\bsnm{Budich}, \binits{J.C.}},
\bauthor{\bsnm{Kunst}, \binits{F.K.}}:
\batitle{Exceptional topology of non-hermitian systems}.
\bjtitle{Rev. Mod. Phys.}
\bvolume{93},
\bfpage{015005}
(\byear{2021})
\doiurl{10.1103/RevModPhys.93.015005}
\end{barticle}
\endbibitem

%%% 52
\bibitem[\protect\citeauthoryear{Hu et~al.}{2022}]{hu-etal.2022}
\begin{barticle}
\bauthor{\bsnm{Hu}, \binits{H.}},
\bauthor{\bsnm{Sun}, \binits{S.}},
\bauthor{\bsnm{Chen}, \binits{S.}}:
\batitle{Knot topology of exceptional point and non-hermitian no-go theorem}.
\bjtitle{Phys. Rev. Res.}
\bvolume{4},
\bfpage{022064}
(\byear{2022})
\doiurl{10.1103/PhysRevResearch.4.L022064}
\end{barticle}
\endbibitem

%%% 53
\bibitem[\protect\citeauthoryear{Yokomizo and Murakami}{2020}]{yokomizo-etal.2020}
\begin{barticle}
\bauthor{\bsnm{Yokomizo}, \binits{K.}},
\bauthor{\bsnm{Murakami}, \binits{S.}}:
\batitle{Topological semimetal phase with exceptional points in one-dimensional non-hermitian systems}.
\bjtitle{Phys. Rev. Res.}
\bvolume{2},
\bfpage{043045}
(\byear{2020})
\end{barticle}
\endbibitem

%%% 54
\bibitem[\protect\citeauthoryear{Ding et~al.}{2022}]{ding-etal.2022}
\begin{barticle}
\bauthor{\bsnm{Ding}, \binits{K.}},
\bauthor{\bsnm{Fang}, \binits{C.}},
\bauthor{\bsnm{Ma}, \binits{G.}}:
\batitle{{Non-Hermitian topology and exceptional-point geometries}}.
\bjtitle{Nature Reviews Physics.}
\bvolume{4},
\bfpage{745}--\blpage{760}
(\byear{2022})
\end{barticle}
\endbibitem

%%% 55
\bibitem[\protect\citeauthoryear{Jia et~al.}{2023}]{jia-etal.2023}
\begin{barticle}
\bauthor{\bsnm{Jia}, \binits{H.}},
\bauthor{\bsnm{Zhang}, \binits{R.-Y.}},
\bauthor{\bsnm{Hu}, \binits{J.}},
\bauthor{\bsnm{Xiao}, \binits{Y.}},
\bauthor{\bsnm{Zhang}, \binits{S.}},
\bauthor{\bsnm{Zhu}, \binits{Y.}},
\bauthor{\bsnm{Chan}, \binits{C.}}:
\batitle{Topological classification for intersection singularities of exceptional surfaces in pseudo-hermitian systems}.
\bjtitle{Communications Physics}
\bvolume{6}(\bissue{1}),
\bfpage{293}
(\byear{2023})
\end{barticle}
\endbibitem

%%% 56
\bibitem[\protect\citeauthoryear{Lai et~al.}{2024}]{lai-etal.2024}
\begin{barticle}
\bauthor{\bsnm{Lai}, \binits{D.-G.}},
\bauthor{\bsnm{Miranowicz}, \binits{A.}},
\bauthor{\bsnm{Nori}, \binits{F.}}:
\batitle{Nonreciprocal topological phonon transfer independent of both device mass and exceptional-point encircling direction}.
\bjtitle{Phys. Rev. Lett.}
\bvolume{132},
\bfpage{243602}
(\byear{2024})
\end{barticle}
\endbibitem

%%% 57
\bibitem[\protect\citeauthoryear{Suntharalingam et~al.}{2023}]{suntharalingam-etal.2023}
\begin{barticle}
\bauthor{\bsnm{Suntharalingam}, \binits{A.}},
\bauthor{\bsnm{Fern{\'a}ndez-Alc{\'a}zar}, \binits{L.}},
\bauthor{\bsnm{Kononchuk}, \binits{R.}},
\bauthor{\bsnm{Kottos}, \binits{T.}}:
\batitle{Noise resilient exceptional-point voltmeters enabled by oscillation quenching phenomena}.
\bjtitle{Nature Communications}
\bvolume{14}(\bissue{1}),
\bfpage{5515}
(\byear{2023})
\end{barticle}
\endbibitem

%%% 58
\bibitem[\protect\citeauthoryear{Felski and Kunst}{2025}]{felski-etal.2025}
\begin{barticle}
\bauthor{\bsnm{Felski}, \binits{A.}},
\bauthor{\bsnm{Kunst}, \binits{F.K.}}:
\batitle{Exceptional points and stability in nonlinear models of population dynamics having $\mathcal{PT}$ symmetry}.
\bjtitle{Phys. Rev. Res.}
\bvolume{7},
\bfpage{013326}
(\byear{2025})
\doiurl{10.1103/PhysRevResearch.7.013326}
\end{barticle}
\endbibitem

%%% 59
\bibitem[\protect\citeauthoryear{Ramezanpour and Bogdanov}{2021}]{PhysRevA.103.043510}
\begin{barticle}
\bauthor{\bsnm{Ramezanpour}, \binits{S.}},
\bauthor{\bsnm{Bogdanov}, \binits{A.}}:
\batitle{Tuning exceptional points with kerr nonlinearity}.
\bjtitle{Phys. Rev. A}
\bvolume{103},
\bfpage{043510}
(\byear{2021})
\doiurl{10.1103/PhysRevA.103.043510}
\end{barticle}
\endbibitem

%%% 60
\bibitem[\protect\citeauthoryear{Ji et~al.}{2023}]{el2023tracking}
\begin{barticle}
\bauthor{\bsnm{Ji}, \binits{K.}},
\bauthor{\bsnm{Zhong}, \binits{Q.}},
\bauthor{\bsnm{Ge}, \binits{L.}},
\bauthor{\bsnm{Beaudoin}, \binits{G.}},
\bauthor{\bsnm{Sagnes}, \binits{I.}},
\bauthor{\bsnm{Raineri}, \binits{F.}},
\bauthor{\bsnm{El-Ganainy}, \binits{R.}},
\bauthor{\bsnm{Yacomotti}, \binits{A.M.}}:
\batitle{Tracking exceptional points above laser threshold}.
\bjtitle{Nat. Commun.}
\bvolume{14},
\bfpage{8304}
(\byear{2023})
\end{barticle}
\endbibitem

%%% 61
\bibitem[\protect\citeauthoryear{Ramezanpour}{2024}]{Ramezanpour:24}
\begin{barticle}
\bauthor{\bsnm{Ramezanpour}, \binits{S.}}:
\batitle{Dynamic of time-independent and time-dependent asymmetric gross-pitaevskii equation around exceptional point}.
\bjtitle{Opt. Continuum}
\bvolume{3},
\bfpage{1907}--\blpage{1917}
(\byear{2024})
\doiurl{10.1364/OPTCON.535433}
\end{barticle}
\endbibitem

%%% 62
\bibitem[\protect\citeauthoryear{Wingenbach et~al.}{2024}]{wingenbach-etal.2024}
\begin{barticle}
\bauthor{\bsnm{Wingenbach}, \binits{J.}},
\bauthor{\bsnm{Schumacher}, \binits{S.}},
\bauthor{\bsnm{Ma}, \binits{X.}}:
\batitle{Manipulating spectral topology and exceptional points by nonlinearity in non-hermitian polariton systems}.
\bjtitle{Phys. Rev. Res.}
\bvolume{6},
\bfpage{013148}
(\byear{2024})
\doiurl{10.1103/PhysRevResearch.6.013148}
\end{barticle}
\endbibitem

%%% 63
\bibitem[\protect\citeauthoryear{Wang et~al.}{2019}]{wang2019dynamics}
\begin{barticle}
\bauthor{\bsnm{Wang}, \binits{H.}},
\bauthor{\bsnm{Assawaworrarit}, \binits{S.}},
\bauthor{\bsnm{Fan}, \binits{S.}}:
\batitle{Dynamics for encircling an exceptional point in a nonlinear non-hermitian system}.
\bjtitle{Optics letters}
\bvolume{44},
\bfpage{638}--\blpage{641}
(\byear{2019})
\end{barticle}
\endbibitem

%%% 64
\bibitem[\protect\citeauthoryear{Li et~al.}{2024}]{Li2024}
\begin{barticle}
\bauthor{\bsnm{Li}, \binits{H.}},
\bauthor{\bsnm{Chen}, \binits{L.}},
\bauthor{\bsnm{Wu}, \binits{W.}},
\bauthor{\bsnm{Wang}, \binits{H.}},
\bauthor{\bsnm{Wang}, \binits{T.}},
\bauthor{\bsnm{Zhong}, \binits{Y.}},
\bauthor{\bsnm{Huang}, \binits{F.}},
\bauthor{\bsnm{Liu}, \binits{G.-S.}},
\bauthor{\bsnm{Chen}, \binits{Y.}},
\bauthor{\bsnm{Luo}, \binits{Y.}},
\bauthor{\bsnm{Chen}, \binits{Z.}}:
\batitle{Enhanced sensitivity with nonlinearity-induced exceptional points degeneracy lifting}.
\bjtitle{Communications Physics}
\bvolume{7},
\bfpage{117}
(\byear{2024})
\doiurl{10.1038/s42005-024-01609-6}
\end{barticle}
\endbibitem

%%% 65
\bibitem[\protect\citeauthoryear{Nikzamir and Capolino}{2022}]{PhysRevApplied.18.054059}
\begin{barticle}
\bauthor{\bsnm{Nikzamir}, \binits{A.}},
\bauthor{\bsnm{Capolino}, \binits{F.}}:
\batitle{Highly sensitive coupled oscillator based on an exceptional point of degeneracy and nonlinearity}.
\bjtitle{Phys. Rev. Appl.}
\bvolume{18},
\bfpage{054059}
(\byear{2022})
\doiurl{10.1103/PhysRevApplied.18.054059}
\end{barticle}
\endbibitem

%%% 66
\bibitem[\protect\citeauthoryear{Bai et~al.}{2023}]{Bai-etal.2023}
\begin{barticle}
\bauthor{\bsnm{Bai}, \binits{K.}},
\bauthor{\bsnm{Fiang}, \binits{L.}},
\bauthor{\bsnm{Liu}, \binits{T.R.}},
\bauthor{\bsnm{Li}, \binits{J.Z.}},
\bauthor{\bsnm{Wan}, \binits{D.}},
\bauthor{\bsnm{Xiao}, \binits{M.}}:
\batitle{{Nonlinearity-enabled higher-order exceptional singularities with ultra-enhanced signal-to-noise ratio}}.
\bjtitle{National Science Review}
\bvolume{10},
\bfpage{259}
(\byear{2023})
\end{barticle}
\endbibitem

%%% 67
\bibitem[\protect\citeauthoryear{Bai et~al.}{2024}]{bai-etal.2024}
\begin{barticle}
\bauthor{\bsnm{Bai}, \binits{K.}},
\bauthor{\bsnm{Liu}, \binits{T.-R.}},
\bauthor{\bsnm{Fang}, \binits{L.}},
\bauthor{\bsnm{Li}, \binits{J.-Z.}},
\bauthor{\bsnm{Lin}, \binits{C.}},
\bauthor{\bsnm{Wan}, \binits{D.}},
\bauthor{\bsnm{Xiao}, \binits{M.}}:
\batitle{Observation of nonlinear exceptional points with a complete basis in dynamics}.
\bjtitle{Phys. Rev. Lett.}
\bvolume{132}(\bissue{7}),
\bfpage{073802}
(\byear{2024})
\end{barticle}
\endbibitem

%%% 68
\bibitem[\protect\citeauthoryear{Isobe et~al.}{2024}]{isobe-etal.2024}
\begin{barticle}
\bauthor{\bsnm{Isobe}, \binits{T.}},
\bauthor{\bsnm{Yoshida}, \binits{T.}},
\bauthor{\bsnm{Hatsugai}, \binits{Y.}}:
\batitle{Bulk-edge correspondence for nonlinear eigenvalue problems}.
\bjtitle{Phys. Rev. Lett.}
\bvolume{132},
\bfpage{126601}
(\byear{2024})
\doiurl{10.1103/PhysRevLett.132.126601}
\end{barticle}
\endbibitem

%%% 69
\bibitem[\protect\citeauthoryear{Sone et~al.}{2022}]{sone-etal.2022}
\begin{barticle}
\bauthor{\bsnm{Sone}, \binits{K.}},
\bauthor{\bsnm{Ashida}, \binits{Y.}},
\bauthor{\bsnm{Sagawa}, \binits{T.}}:
\batitle{Topological synchronization of coupled nonlinear oscillators}.
\bjtitle{Phys. Rev. Res.}
\bvolume{4},
\bfpage{023211}
(\byear{2022})
\doiurl{10.1103/PhysRevResearch.4.023211}
\end{barticle}
\endbibitem

%%% 70
\bibitem[\protect\citeauthoryear{Bender}{2017}]{bender.2017}
\begin{bchapter}
\bauthor{\bsnm{Bender}, \binits{C.M.}}:
\bctitle{Nonlinear eigenvalue problems and pt-symmetric quantum mechanics}.
In: \bbtitle{Journal of Physics: Conference Series},
vol. \bseriesno{873},
p. \bfpage{012002}
(\byear{2017}).
\bcomment{IOP Publishing}
\end{bchapter}
\endbibitem

%%% 71
\bibitem[\protect\citeauthoryear{Zeeman}{1976}]{zeeman.1976b}
\begin{barticle}
\bauthor{\bsnm{Zeeman}, \binits{E.C.}}:
\batitle{{Catastrophe Theory}}.
\bjtitle{{Scientific American Magazine}}
\bvolume{234},
\bfpage{65}--\blpage{83}
(\byear{1976})
\end{barticle}
\endbibitem

%%% 72
\bibitem[\protect\citeauthoryear{Thom}{1989}]{thom.1989}
\begin{bbook}
\bauthor{\bsnm{Thom}, \binits{R.}}:
\bbtitle{{S}tructural {S}tability of {M}orphogenesis}.
\bpublisher{{CRC Press}},
\blocation{Reading}
(\byear{1989})
\end{bbook}
\endbibitem

%%% 73
\bibitem[\protect\citeauthoryear{Arnold}{1992}]{arnold.1992}
\begin{bbook}
\bauthor{\bsnm{Arnold}, \binits{V.}}:
\bbtitle{{C}atastrophe {T}heory, 3rd ed}.
\bpublisher{Springer},
\blocation{Berlin}
(\byear{1992})
\end{bbook}
\endbibitem

%%% 74
\bibitem[\protect\citeauthoryear{Saunders}{1980}]{saunders.80}
\begin{bbook}
\bauthor{\bsnm{Saunders}, \binits{P.T.}}:
\bbtitle{An Introduction to Catastrophe Theory}.
\bpublisher{Cambridge University Press},
\blocation{Cambridge}
(\byear{1980})
\end{bbook}
\endbibitem

%%% 75
\bibitem[\protect\citeauthoryear{Gilmore}{1981}]{gilmore.81}
\begin{bbook}
\bauthor{\bsnm{Gilmore}, \binits{R.}}:
\bbtitle{Catastrophe Theory for Scientists and Engineers}.
\bpublisher{Wiley},
\blocation{New York}
(\byear{1981})
\end{bbook}
\endbibitem

%%% 76
\bibitem[\protect\citeauthoryear{Poston and Stewart}{1978}]{poston-stewart.1978}
\begin{bbook}
\bauthor{\bsnm{Poston}, \binits{T.}},
\bauthor{\bsnm{Stewart}, \binits{I.}}:
\bbtitle{{C}atastrophe: {T}heory and {I}ts {A}pplications}.
\bpublisher{Dover},
\blocation{{New York}}
(\byear{1978})
\end{bbook}
\endbibitem

%%% 77
\bibitem[\protect\citeauthoryear{{H. A. Adam}}{2006}]{adam.2002}
\begin{barticle}
\bauthor{\bsnm{{H. A. Adam}}}:
\batitle{{The mathematical physics of rainbows and glories}}.
\bjtitle{Physics Reports}
\bvolume{548},
\bfpage{229}--\blpage{365}
(\byear{2006})
\end{barticle}
\endbibitem

%%% 78
\bibitem[\protect\citeauthoryear{Berry and Upstill}{1980}]{berry-upstill.1980}
\begin{barticle}
\bauthor{\bsnm{Berry}, \binits{M.V.}},
\bauthor{\bsnm{Upstill}, \binits{C.}}:
\batitle{{Catastrophe optics: morphology of caustics and their diffraction patterns}}.
\bjtitle{Progress in Optics}
\bvolume{{XVIII}},
\bfpage{259}--\blpage{343}
(\byear{1980})
\end{barticle}
\endbibitem

%%% 79
\bibitem[\protect\citeauthoryear{Berry}{1990}]{berry.90}
\begin{barticle}
\bauthor{\bsnm{Berry}, \binits{M.}}:
\batitle{Beyond rainbows}.
\bjtitle{Current Science}
\bvolume{59},
\bfpage{1175}--\blpage{1182}
(\byear{1990})
\end{barticle}
\endbibitem

%%% 80
\bibitem[\protect\citeauthoryear{Berry et~al.}{1979}]{berry-etal.79}
\begin{barticle}
\bauthor{\bsnm{Berry}, \binits{M.V.}},
\bauthor{\bsnm{Nye}, \binits{J.F.}},
\bauthor{\bsnm{Wright}, \binits{F.J.}}:
\batitle{The elliptic umbilic diffraction catastrophe}.
\bjtitle{Philosophical Transcations of the Royal Society of London A: Mathematical, Physical and Engineering Sciences}
\bvolume{291},
\bfpage{32}--\blpage{63}
(\byear{1979})
\end{barticle}
\endbibitem

%%% 81
\bibitem[\protect\citeauthoryear{Zeeman}{1976}]{zeeman.1976a}
\begin{bchapter}
\bauthor{\bsnm{Zeeman}, \binits{E.C.}}:
\bctitle{Brain modelling}.
In: \beditor{\bsnm{Hilton}, \binits{P.}} (ed.)
\bbtitle{Structural Stability, the Theory of Catastrophes, and Applications in the Sciences},
pp. \bfpage{367}--\blpage{372}.
\bpublisher{Springer},
\blocation{Berlin, Heidelberg}
(\byear{1976})
\end{bchapter}
\endbibitem

%%% 82
\bibitem[\protect\citeauthoryear{{R. Rosen}}{1988}]{rosen.1988}
\begin{barticle}
\bauthor{\bsnm{{R. Rosen}}}:
\batitle{{How universal is a universal unfolding?}}
\bjtitle{{Appl. Math. Lett.}}
\bvolume{{1}},
\bfpage{105}--\blpage{107}
(\byear{1988})
\end{barticle}
\endbibitem

%%% 83
\bibitem[\protect\citeauthoryear{Heiss}{2004}]{heiss.04}
\begin{barticle}
\bauthor{\bsnm{Heiss}, \binits{W.D.}}:
\batitle{Exceptional points of non-hermitian operators}.
\bjtitle{Journal of Physics: Mathematical and General}
\bvolume{37},
\bfpage{2455}
(\byear{2004})
\end{barticle}
\endbibitem

%%% 84
\bibitem[\protect\citeauthoryear{Barkley-Rosser}{2007}]{barkleyrosser.2007}
\begin{barticle}
\bauthor{\bsnm{Barkley-Rosser}, \binits{J.}}:
\batitle{{The rise and fall of catastrophe theory applications in economics: Was the baby thrown out with the bathwater?}}
\bjtitle{Journal of Economic Dynamics and Control.}
\bvolume{31},
\bfpage{3255}--\blpage{3280}
(\byear{2007})
\end{barticle}
\endbibitem

%%% 85
\bibitem[\protect\citeauthoryear{Brocker and Lander}{1975}]{broecker-lander.1975}
\begin{bbook}
\bauthor{\bsnm{Brocker}, \binits{T.}},
\bauthor{\bsnm{Lander}, \binits{L.}}:
\bbtitle{{D}ifferentiable {G}erms and {C}atastrophes}.
\bpublisher{{Cambridge University Press}},
\blocation{London}
(\byear{1975})
\end{bbook}
\endbibitem

%%% 86
\bibitem[\protect\citeauthoryear{Liu et~al.}{2000}]{Liu2000}
\begin{barticle}
\bauthor{\bsnm{Liu}, \binits{W.-M.}},
\bauthor{\bsnm{Wu}, \binits{B.}},
\bauthor{\bsnm{Niu}, \binits{Q.}}:
\batitle{Nonlinear effects in interference of {B}ose-{E}instein condensates}.
\bjtitle{Phys. Rev. Lett.}
\bvolume{84},
\bfpage{2294}
(\byear{2000})
\end{barticle}
\endbibitem

%%% 87
\bibitem[\protect\citeauthoryear{Berloff}{1999}]{berloff1999}
\begin{barticle}
\bauthor{\bsnm{Berloff}, \binits{N.G.}}:
\batitle{Nonlocal nonlinear {S}chr\"odinger equations as models of superfluidity}.
\bjtitle{Journal of Low Temperature Physics}
\bvolume{116},
\bfpage{359}--\blpage{380}
(\byear{1999})
\end{barticle}
\endbibitem

%%% 88
\bibitem[\protect\citeauthoryear{Kivshar and Malomed}{1989}]{Kivshar1989}
\begin{barticle}
\bauthor{\bsnm{Kivshar}, \binits{Y.S.}},
\bauthor{\bsnm{Malomed}, \binits{B.A.}}:
\batitle{Dynamics of solitons in nearly integrable systems}.
\bjtitle{Rev. Mod. Phys.}
\bvolume{61},
\bfpage{763}
(\byear{1989})
\end{barticle}
\endbibitem

%%% 89
\bibitem[\protect\citeauthoryear{Dominici et~al.}{2015}]{Dominici2015}
\begin{barticle}
\bauthor{\bsnm{Dominici}, \binits{L.}},
\bauthor{\bsnm{Dagvadorj}, \binits{G.}},
\bauthor{\bsnm{Fellows}, \binits{J.M.}},
\bauthor{\bsnm{Ballarini}, \binits{D.}},
\bauthor{\bsnm{{De Giorgi}}, \binits{M.}},
\bauthor{\bsnm{Marchetti}, \binits{F.M.}},
\bauthor{\bsnm{Piccirillo}, \binits{B.}},
\bauthor{\bsnm{Marrucci}, \binits{L.}},
\bauthor{\bsnm{Bramati}, \binits{A.}},
\bauthor{\bsnm{Gigli}, \binits{G.}},
\bauthor{\bsnm{Szymanska}, \binits{M.H.}},
\bauthor{\bsnm{Sanvitto}, \binits{D.}}:
\batitle{Vortex and half-vortex dynamics in a nonlinear spinor quantum fluid}.
\bjtitle{Science Adv.}
\bvolume{1},
\bfpage{1500807}
(\byear{2015})
\end{barticle}
\endbibitem

%%% 90
\bibitem[\protect\citeauthoryear{Saleh and Teich}{2006}]{saleh-teich.06}
\begin{bbook}
\bauthor{\bsnm{Saleh}, \binits{B.E.A.}},
\bauthor{\bsnm{Teich}, \binits{M.C.}}:
\bbtitle{Fundamentals of Photonics},
\bedition{{2nd}} edn.
\bpublisher{Wiley},
\blocation{New York}
(\byear{2006})
\end{bbook}
\endbibitem

%%% 91
\bibitem[\protect\citeauthoryear{Martijn~de Sterke and Sipe}{1990}]{sterke-sipe.1990}
\begin{barticle}
\bauthor{\bsnm{Sterke}, \binits{C.}},
\bauthor{\bsnm{Sipe}, \binits{J.E.}}:
\batitle{Coupled modes and the nonlinear schr\"odinger equation}.
\bjtitle{Phys. Rev. A}
\bvolume{42},
\bfpage{550}--\blpage{555}
(\byear{1990})
\doiurl{10.1103/PhysRevA.42.550}
\end{barticle}
\endbibitem

%%% 92
\bibitem[\protect\citeauthoryear{Scott}{2003}]{scott.2003}
\begin{bbook}
\bauthor{\bsnm{Scott}, \binits{A.}}:
\bbtitle{Nonlinear Science},
\bedition{2nd} edn.
\bpublisher{Oxford University Press},
\blocation{New York}
(\byear{2003})
\end{bbook}
\endbibitem

%%% 93
\bibitem[\protect\citeauthoryear{Cruzeiro-Hansson et~al.}{1988}]{cruzeirohansson-etal.1988}
\begin{barticle}
\bauthor{\bsnm{Cruzeiro-Hansson}, \binits{L.}},
\bauthor{\bsnm{Christiansen}, \binits{P.L.}},
\bauthor{\bsnm{Elgin}, \binits{J.N.}}:
\batitle{Comment on ``self-trapping on a dimer: Time-dependent solutions of a discrete nonlinear schr\"odinger equation''}.
\bjtitle{Phys. Rev. B}
\bvolume{37},
\bfpage{7896}--\blpage{7897}
(\byear{1988})
\doiurl{10.1103/PhysRevB.37.7896}
\end{barticle}
\endbibitem

%%% 94
\bibitem[\protect\citeauthoryear{Delplace et~al.}{2021}]{delplace-etal.2021}
\begin{barticle}
\bauthor{\bsnm{Delplace}, \binits{P.}},
\bauthor{\bsnm{Yoshida}, \binits{T.}},
\bauthor{\bsnm{Hatsugai}, \binits{Y.}}:
\batitle{Symmetry-protected multifold exceptional points and their topological characterization}.
\bjtitle{Phys. Rev. Lett.}
\bvolume{127},
\bfpage{186602}
(\byear{2021})
\end{barticle}
\endbibitem

%%% 95
\bibitem[\protect\citeauthoryear{Hu et~al.}{2023}]{hu-etal.2023}
\begin{barticle}
\bauthor{\bsnm{Hu}, \binits{J.}},
\bauthor{\bsnm{Zhang}, \binits{R.-Y.}},
\bauthor{\bsnm{Wang}, \binits{Y.}},
\bauthor{\bsnm{Ouyang}, \binits{X.}},
\bauthor{\bsnm{Zhu}, \binits{Y.}},
\bauthor{\bsnm{Jia}, \binits{H.}},
\bauthor{\bsnm{Chan}, \binits{C.T.}}:
\batitle{Non-hermitian swallowtail catastrophe revealing transitions among diverse topological singularities}.
\bjtitle{Nature Physics}
\bvolume{19}(\bissue{8}),
\bfpage{1098}--\blpage{1103}
(\byear{2023})
\end{barticle}
\endbibitem

%%% 96
\bibitem[\protect\citeauthoryear{Panahi et~al.}{2024}]{panahi-etal.2024}
\begin{barticle}
\bauthor{\bsnm{Panahi}, \binits{S.}},
\bauthor{\bsnm{Ye}, \binits{L.-L.}},
\bauthor{\bsnm{Lai}, \binits{Y.-C.}}:
\batitle{Higher-order exceptional points and stochastic resonance in pseudo-hermitian systems}.
\bjtitle{Phys. Rev. Appl.}
\bvolume{22},
\bfpage{054063}
(\byear{2024})
\doiurl{10.1103/PhysRevApplied.22.054063}
\end{barticle}
\endbibitem

%%% 97
\bibitem[\protect\citeauthoryear{Minganti et~al.}{2019}]{minganti-etal.2019}
\begin{barticle}
\bauthor{\bsnm{Minganti}, \binits{F.}},
\bauthor{\bsnm{Miranowicz}, \binits{A.}},
\bauthor{\bsnm{Chhajlany}, \binits{R.W.}},
\bauthor{\bsnm{Nori}, \binits{F.}}:
\batitle{Quantum exceptional points of non-hermitian hamiltonians and liouvillians: The effects of quantum jumps}.
\bjtitle{Phys. Rev. A}
\bvolume{100},
\bfpage{062131}
(\byear{2019})
\doiurl{10.1103/PhysRevA.100.062131}
\end{barticle}
\endbibitem

%%% 98
\bibitem[\protect\citeauthoryear{Sun and Yi}{2024}]{sun-yi.2024}
\begin{barticle}
\bauthor{\bsnm{Sun}, \binits{K.}},
\bauthor{\bsnm{Yi}, \binits{W.}}:
\batitle{Encircling the liouvillian exceptional points: a brief review}.
\bjtitle{AAPPS Bulletin}
\bvolume{34}(\bissue{1}),
\bfpage{22}
(\byear{2024})
\end{barticle}
\endbibitem

%%% 99
\bibitem[\protect\citeauthoryear{Castrigiano and Hayes}{2004}]{castrigiano-hayes.2004}
\begin{bbook}
\bauthor{\bsnm{Castrigiano}, \binits{D.P.L.}},
\bauthor{\bsnm{Hayes}, \binits{S.A.}}:
\bbtitle{Catastrophe Theory, 2nd Ed.}
\bpublisher{Westview Press},
\blocation{Boulder}
(\byear{2004})
\end{bbook}
\endbibitem

\end{thebibliography}

\vspace*{1cm}


\vspace*{1cm}

%\begin{acknowledgments}
{\bf Acknowledgments} \\
    The authors gratefully acknowledge financial support for the Arizona group from the US National Science Foundation (NSF) under Grant No. DMR-1839570, and, for  the Paderborn groups, by the Deutsche Forschungsgemeinschaft (DFG, German Research Foundation) through Grant No.~467358803 and the transregional collaborative research center TRR 142 (Projects A04 and C10, Grant No. 231447078). We acknowledge support for the publication cost by the Open Access Publication Fund of Paderborn University.
%\end{acknowledgments}

\vspace*{1cm}

{\bf Author Contributions} \\
N.H.K, J.W., S.S. and R.B. conceived the project,
N.H.K. developed the bifurcation theory, J.W. provided numerical solutions, L.A., J.S. and X.M. provided the solution algorithm for the nonlinear problem, R.B. and S.S. wrote the paper with input from all authors.

\vspace*{1cm}

{\bf Competing interests} \\
The authors declare no competing interests.

%=======================================

%\end{widetext}

\end{document}